\documentclass[final,1p,times]{nim/elsarticle}
\usepackage{epsfig,textcomp,subfigure,units}

\newcommand{\bal}{\begin{eqnarray*}}
\newcommand{\eal}{\end{eqnarray*}}
\def\apname{}

\journal{Nuclear Inst. and Methods in Physics Research, A}

\begin{document}

\begin{frontmatter}

\title{Measurement of South Pole ice transparency with the IceCube LED calibration system}

\author[Adelaide]{M.~G.~Aartsen}
\author[MadisonPAC]{R.~Abbasi}
\author[Gent]{Y.~Abdou}
\author[Zeuthen]{M.~Ackermann}
\author[Christchurch]{J.~Adams}
\author[Geneva]{J.~A.~Aguilar}
\author[MadisonPAC]{M.~Ahlers}
\author[Berlin]{D.~Altmann}
\author[MadisonPAC]{J.~Auffenberg}
\author[Bartol]{X.~Bai\fnref{SouthDakota}}
% ...
\address[Adelaide]{School of Chemistry \& Physics, University of Adelaide, Adelaide SA, 5005 Australia}
% ...
\address[Berlin]{Institut f\"ur Physik, Humboldt-Universit\"at zu Berlin, D-12489 Berlin, Germany}
% ...
\address[Christchurch]{Dept.~of Physics and Astronomy, University of Canterbury, Private Bag 4800, Christchurch, New Zealand}
% ...
\address[Geneva]{D\'epartement de physique nucl\'eaire et corpusculaire, Universit\'e de Gen\`eve, CH-1211 Gen\`eve, Switzerland}
\address[Gent]{Dept.~of Physics and Astronomy, University of Gent, B-9000 Gent, Belgium}
% ...
\address[MadisonPAC]{Dept.~of Physics and Wisconsin IceCube Particle Astrophysics Center, University of Wisconsin, Madison, WI 53706, USA}
% ...
\address[Bartol]{Bartol Research Institute and Department of Physics and Astronomy, University of Delaware, Newark, DE 19716, USA}
% ...
\address[Zeuthen]{DESY, D-15735 Zeuthen, Germany}
\fntext[SouthDakota]{Physics Department, South Dakota School of Mines and Technology, Rapid City, SD 57701, USA}
\fntext[LosAlamos]{Los Alamos National Laboratory, Los Alamos, NM 87545, USA}
\fntext[Bari]{also Sezione INFN, Dipartimento di Fisica, I-70126, Bari, Italy}
\fntext[Goddard]{NASA Goddard Space Flight Center, Greenbelt, MD 20771, USA}

\begin{abstract}
\label{abs}
The IceCube Neutrino Observatory, approximately 1 km$^3$ in size, is now complete with 86 strings deployed in the Antarctic ice.
IceCube detects the Cherenkov radiation emitted by charged particles passing through or created in the ice.
To realize the full potential of the detector, the properties of light propagation in the ice in and around the detector must be well understood.
This report presents a new method of fitting the model of light propagation in the ice to a data set of in-situ light source events collected with IceCube. The resulting set of derived parameters, namely the measured values of scattering and absorption coefficients vs.\ depth, is presented and a comparison of IceCube data with simulations based on the new model is shown.
\end{abstract}

\cortext[cor1]{Corresponding author}
\fntext[fn1]{dima@icecube.wisc.edu}

\end{frontmatter}

\section{Introduction}

\begin{figure}[!h]\begin{center}
\begin{tabular}{c}
\mbox{\epsfig{file=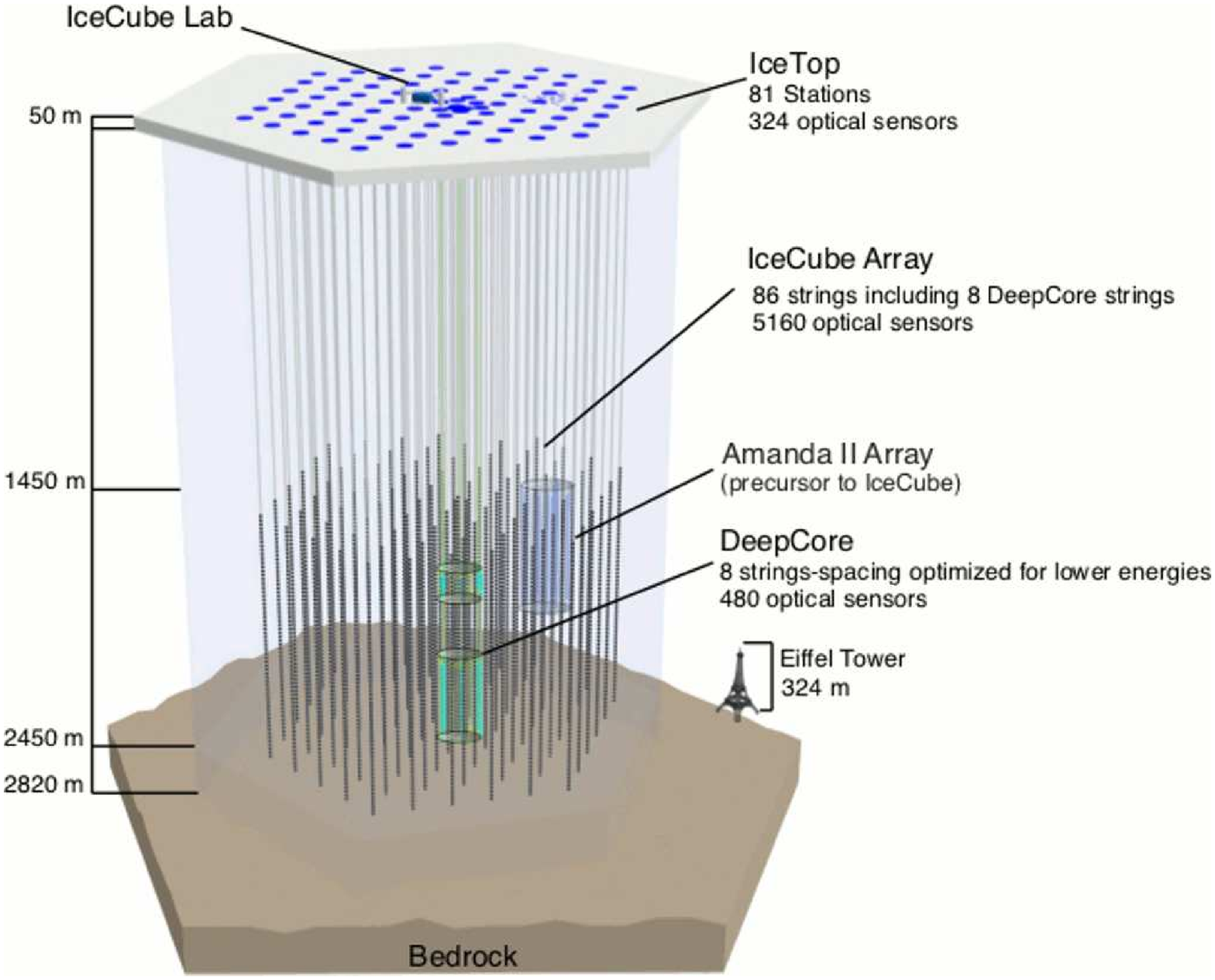,width=.7\textwidth}} \\
\end{tabular}
\parbox{.9\textwidth}{\caption{\label{fullIC} The IceCube Neutrino Observatory, final configuration. Also shown is the AMANDA array, precursor to IceCube, which ended operation in 2009.}}
\end{center}\end{figure}

IceCube is a cubic-kilometer-scale high-energy neutrino observatory built at the geographic South Pole \cite{perf} (see Fig.\ \ref{fullIC}). A primary goal of IceCube is to elucidate the mechanisms for production of high-energy cosmic rays by detecting high-energy neutrinos from astrophysical sources. IceCube uses the 2.8~km thick glacial ice sheet as a medium for producing Cherenkov light emitted by charged particles created when neutrinos interact in the ice or nearby rock. Neutrino interactions can create high-energy muons, electrons or tau leptons, which must be distinguished from a background of downgoing atmospheric muons based on the pattern of emitted Cherenkov light. This light is detected by an embedded array of 5160 optical sensors (digital optical modules, or DOMs for short), 4680 of which are deployed at depths of 1450 - 2450~m and spaced 17~m apart along 78 vertical cables (strings). The strings are arranged in a triangular lattice with a horizontal spacing of approximately 125 m.
The remaining 480 sensors are deployed in a more compact geometry forming the center of the DeepCore array \cite{deepcore}.
The IceCube optical sensors are remotely-controlled autonomous detection units which digitize the data. They include light-emitting diodes (LEDs) which may be used as artificial in-situ light sources.
Also shown in Fig.\ \ref{fullIC} is the location of the AMANDA-II neutrino telescope. AMANDA-II was the precursor for IceCube and was composed of 677 optical sensors organized along 19 strings, with most of the sensors located at depths of 1500 to 2000~m. It operated as a part of the IceCube observatory until it was decommissioned in May 2009.

\begin{figure}[!h]\begin{center}
\begin{tabular}{ccc}
\mbox{\epsfig{file=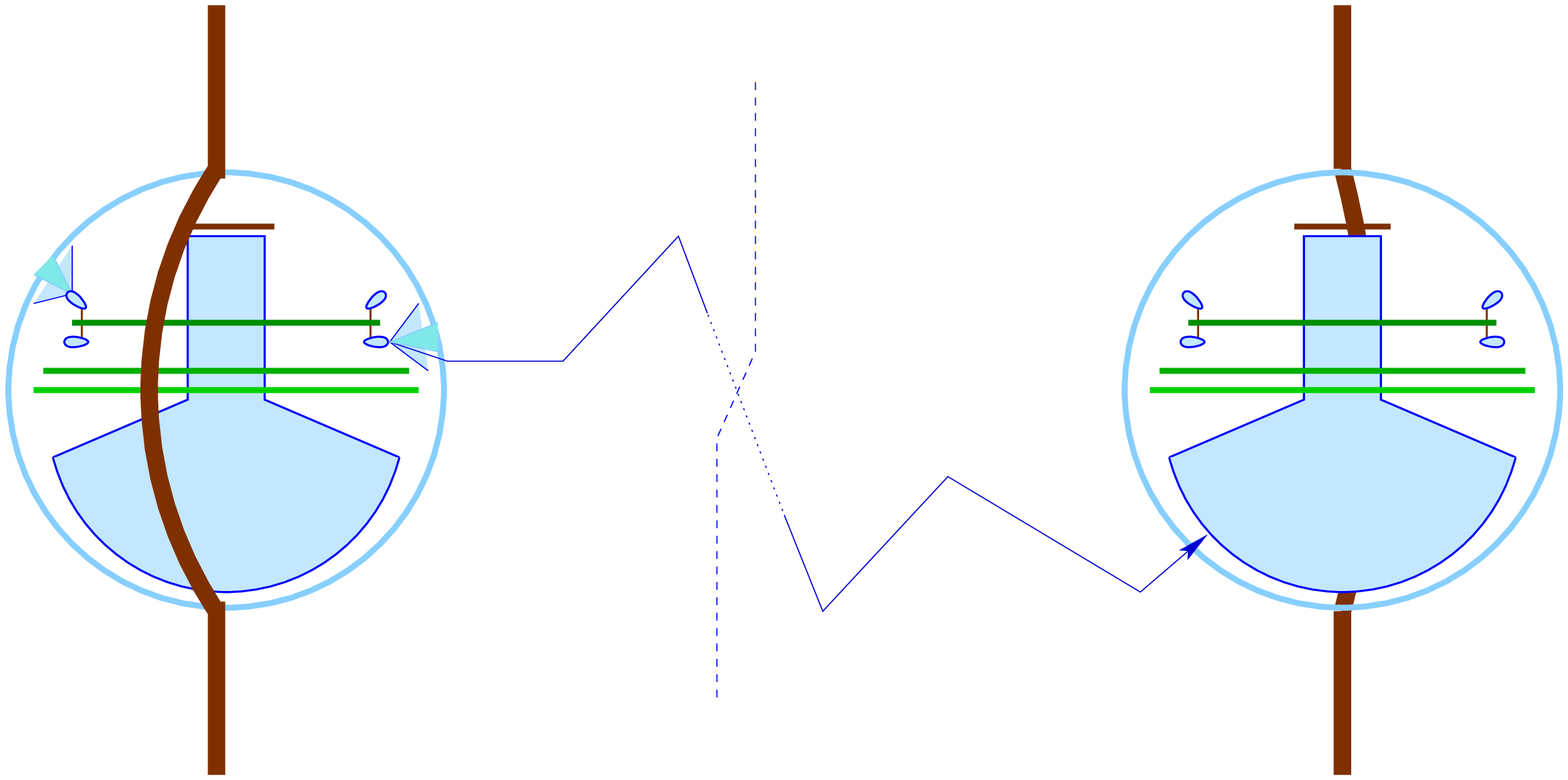,width=.45\textwidth}} & \ & \mbox{\epsfig{file=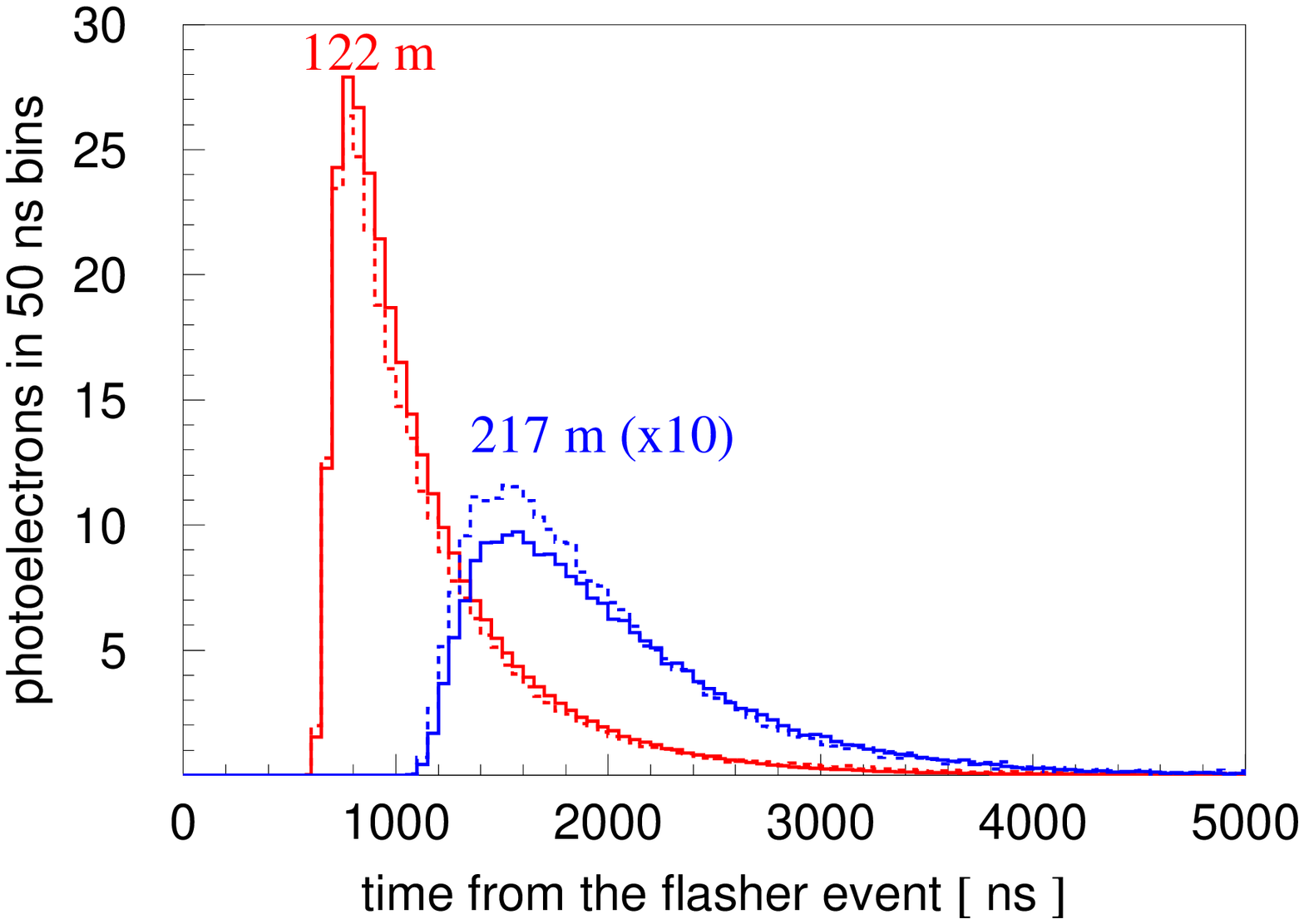,width=.45\textwidth}} \\
\end{tabular}
\parbox{.9\textwidth}{\caption{\label{schema} Left (a): simplified schematics of the experimental setup: the flashing sensor on the left emits photons, which propagate through ice and are detected by a receiving sensor on the right. Right (b): example photon arrival time distributions at a sensor on one of the nearest strings (122~m away) and on one of the next-to-nearest strings (217~m away; histogram values are multiplied by a factor of 10 for clarity). Dashed lines show data and solid lines show simulation based on the model of this work (with best fit parameters). The goal of this work is to find the best-fit ice parameters that describe these distributions as observed in data simultaneously for all pairs of emitters and receivers. }}
\end{center}\end{figure}

Cherenkov photons are emitted with a characteristic wavelength dependence of $1/\lambda^2$ in the wavelength range of 300-600 nm, which includes the relevant sensitivity region of the photosensors. Photons are emitted in a cone around the direction of particle motion with an opening angle, determined by the speed of the particle and refractive index of the ice \cite{rind}, of about $41^\circ$ for relativistic particles.
As the photons propagate from the point of emission to the receiving sensor, they are affected by absorption and scattering in the ice.
These propagation effects must be considered for both simulation and reconstruction of IceCube data and thus need to be carefully modeled.
The important parameters to describe photon propagation in a transparent medium are: the average distance to absorption, the average distance between successive scatters of photons, and the angular distribution of the new direction of a photon at each given scattering point.
This work presents a new, global-fit approach which achieves an improved description of experimental data.
 
To determine the ice parameters, dedicated measurements are performed with the IceCube detector. Photons are emitted by the LEDs in DOMs and recorded by other DOMs, as sketched in Fig.\ \ref{schema}a. The recorded data include the total charge (corresponding to the number of arriving photons) and photon arrival times, shown in Fig.\ \ref{schema}b. A data set that covers all detector depths was produced. A global fit of these data was performed, and the result is a set of scattering and absorption parameters that best describes the full data set.
The AMANDA Collaboration used an analysis based on separate fits to data for individual pairs of emitters and receivers \cite{kurt} to measure the optical properties of the ice. These fits used data taken at very low light levels, to avoid multi-photon pileup detector effects.
%This differs from the approach presented in \cite{kurt} where separate fits were made to data for individual pairs of emitters and receivers in the AMANDA detector, each resulting in a measure of the average properties of the surrounding ice. These properties were then averaged for nearby pairs, resulting in a table of ice parameters.

The relevant detector instrumentation is described in Section \ref{instrumentation} of this paper. Section \ref{flasherdata} introduces the data set. The parameterization for modeling the ice surrounding the detector is described in Section \ref{six}, while Section \ref{simulation} discusses the simulation. The likelihood function used to compare data and simulation is discussed in Sections \ref{llh} and \ref{reg}, and Section \ref{fitting} explains how the search for the best solution was performed. Section \ref{logger} compares the result with an independent probe of the dust concentration in ice \cite{ryan}. Finally, Section \ref{errors} discusses the uncertainties of the measurement, Section \ref{comparison} presents data-simulation comparisons, and Section \ref{results} summarizes the result.

\section{Instrumentation}
\label{instrumentation}

\begin{figure}[!h]\begin{center}
\begin{tabular}{ccc}
\mbox{\epsfig{file=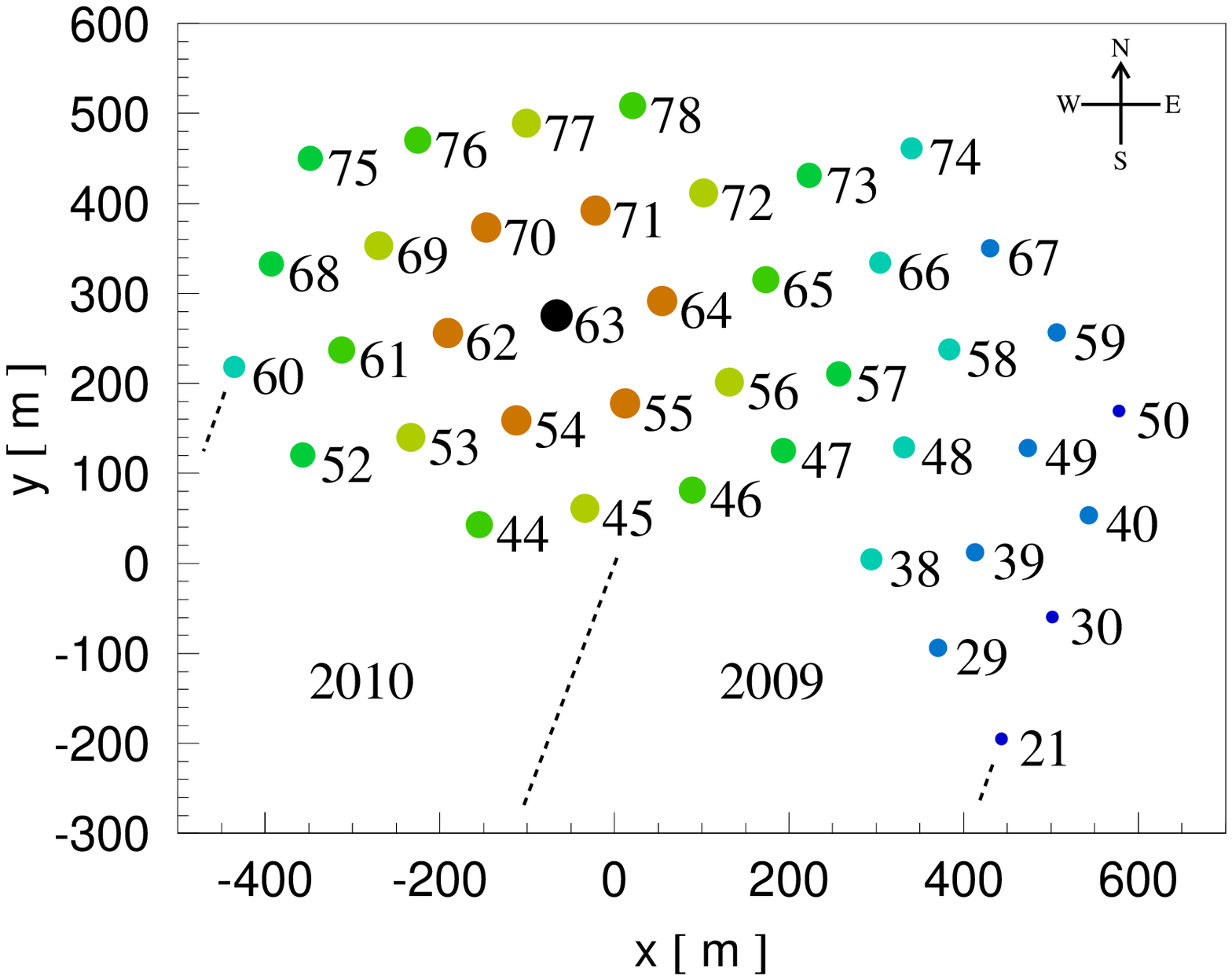,width=.45\textwidth}} & \ & \mbox{\epsfig{file=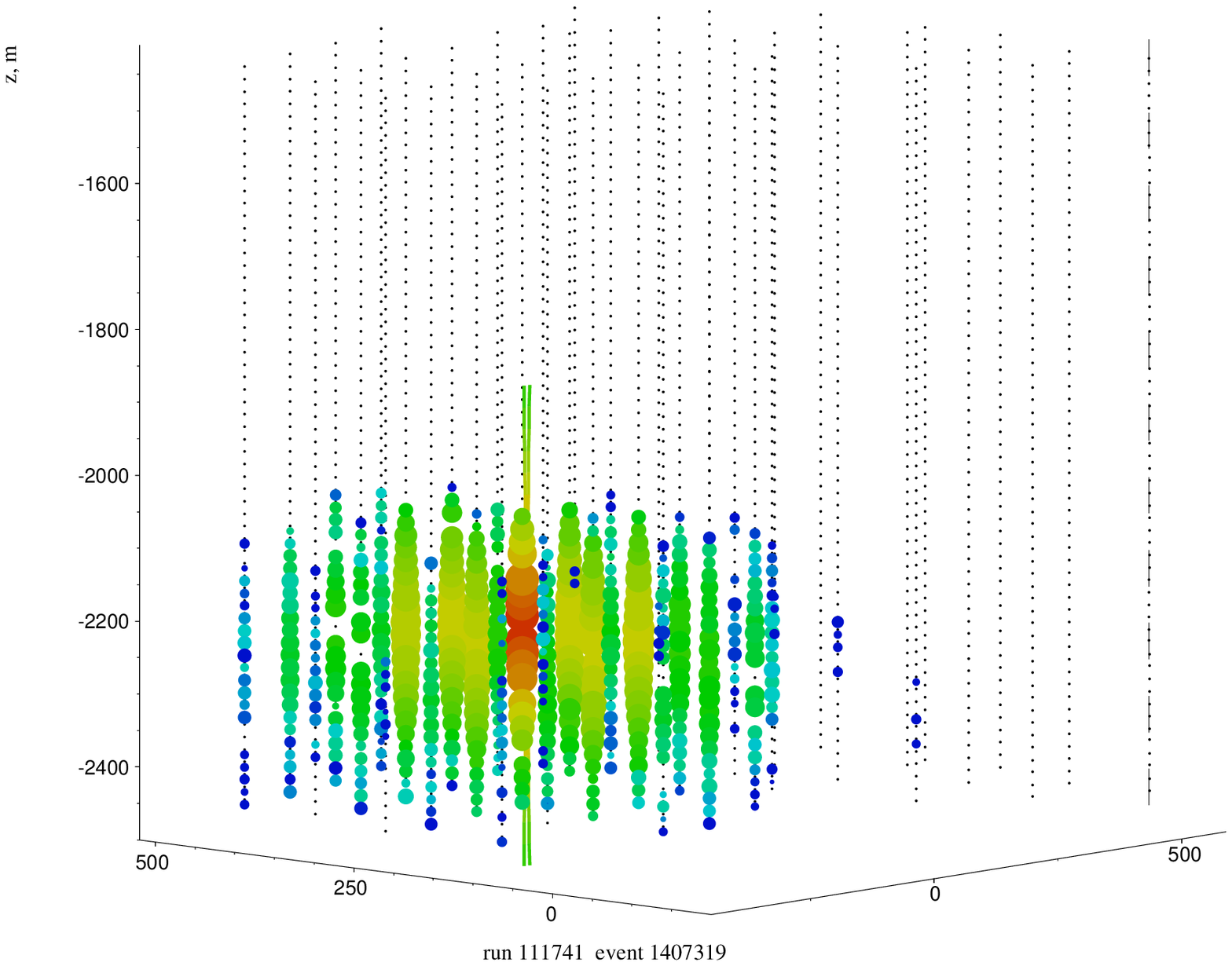,width=.45\textwidth}} \\
\end{tabular}
\parbox{.9\textwidth}{\caption{\label{geo} Left: Top view layout of IceCube in the 40-string configuration in 2008. String 63, for which the DOMs emitted flashing light in the study presented here, is shown in black. The nearest 6 strings are shown in brown. The dashed lines and numbers 2009 and 2010 in the left figure indicate the approximate location of the detector parts deployed during those years. Right: a typical DOM flasher event, DOM 46 on string 63 flashing. The larger circles represent DOMs that recorded larger numbers of photons. The arrival time of the earliest photon in each DOM is indicated with color: early times are shown in red while late times trend to blue.}}
\end{center}\end{figure}

The data for this analysis were obtained in 2008 when IceCube consisted of 40 strings, each with 60 DOMs, as shown in Fig.\ \ref{geo}. Each of the DOMs consists of a 10" 
photomultiplier tube (PMT) \cite{pmt} facing downwards and several electronics boards enclosed in a glass pressure sphere \cite{perf}.  
The main board of the electronics includes two types of digitizers for recording PMT waveforms as well as 
time stamping, control and communications \cite{daq}.  The first 427~ns of each waveform is 
digitized at 300~megasamples per second by fast ATWD chips (analog transient waveform digitizer, see \cite{daq}), and longer duration signals are recorded 
at 25~megasamples per second by the fast ADC (fast analog-to-digital converter, or fADC for short) chips.
The system is capable of resolving charge of up to 300 photoelectrons per 25 ns with precision limited only by the properties of the PMTs ({\it i.e.}, 1 photoelectron is resolved with $\sim 25$\% uncertainty in charge).
Both the ATWD and fADC use 10 bits for amplitude digitization.
However, the ATWD uses three parallel channels with different gains (with a factor of about 8 between) and has a finer time resolution than the fADC (roughly 3.3 vs.\ 25 ns bin width).
The main board contains two ATWD chips on each DOM, ensuring that a waveform can be recorded with one chip while the other one is read out, thus reducing the sensor dead time.

Each DOM includes 12 LEDs on a ``flasher board'' that produce pulsed light detectable by other DOMs located up to 0.5 km away.
The primary purpose of the measurements with these flashers is calibration of the detector. These calibration studies include determining the detector geometry, verifying the calibration of time offsets and the time resolution, verifying the linearity of photon intensity measurement, and extracting the optical properties of the detector ice (this paper).

\begin{figure}[!h]
\begin{center}
\includegraphics[width=.5\textwidth]{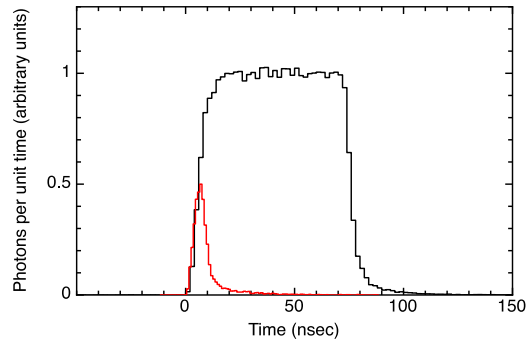}
\caption{
Flasher light output time profile for pulses of minimum and maximum width.  The relative height of the short pulse has been scaled so the leading edges are comparable. This measurement was performed using a small PMT (Hamamatsu R1450) after optical attenuation of the pulses to facilitate counting of individual photons.
}
\label{fig:flasher_pulse_shape}
\end{center}
\end{figure}

Depending on the intended application,
flasher pulses can be programmed with rates from 1.2~Hz to 610~Hz, durations of up to 70~ns,
and LED currents up to 240~mA. 
The corresponding total output from each LED ranges from 
below $10^6$ to about $10^{10}$ photons. 
The programmed current pulse is applied to each individual LED through a high-speed MOSFET (metal-oxide-semiconductor field-effect transistor) driver with a series resistor.  The voltage across the 
resistor is recorded by the DOM's waveform digitizer to precisely define the onset of each pulse.
Figure~\ref{fig:flasher_pulse_shape} shows laboratory measurements 
of the optical output time profiles from short and wide pulses.  
Pulses exhibit a characteristic rise time of 3--4~ns and
a small afterglow, decaying with a 12~ns time constant.  
The narrowest pulses achievable have a full width at half maximum (FWHM) of 6~ns.  

\begin{figure}[!h]
\begin{center}
\includegraphics[width=.5\textwidth]{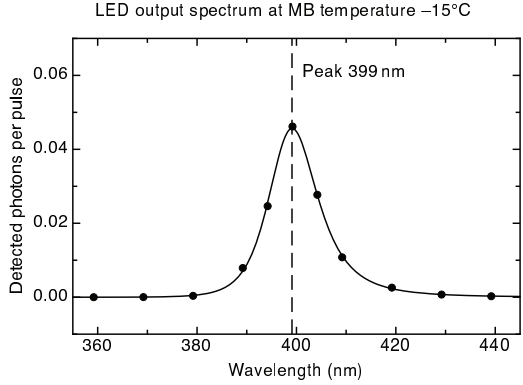}
\caption
{
Wavelength spectrum of light emitted for a DOM operating at a mainboard (MB) temperature of $-15^\circ$ C. The y-axis shows the average number of photons detected per pulse of the LED light.
}
\label{fig:flasher_wavelength_spectrum}
\end{center}
\end{figure}

The wavelength 
spectrum has been measured for the LED light exiting the glass pressure sphere and was found to be centered at 399~nm 
with a FWHM of 14~nm (see Fig.~\ref{fig:flasher_wavelength_spectrum}).  
This wavelength was chosen to approximate
the typical wavelength of detected Cherenkov photons (as discussed and shown in Fig.\ \ref{oms2} below).
To supplement data from the standard flashers, 16 special DOMs 
were constructed and deployed with LEDs that emit at 340~nm, 370~nm, 450~nm, and 500~nm.
Data from these special flashers
were not used in the analysis of this paper but will be used in future analyses of wavelength-dependent effects.

The 12 LEDs in each DOM are aimed in six different azimuth angles (with $60^\circ$ spacing) and 
along two different zenith angles.  After correcting for refraction at interfaces between air, glass and ice, the angular
emission profiles peak along the horizontal direction for the 6 horizontal LEDs and $48^\circ$ above the horizontal for the 6 tilted LEDs.  The angular
spread is reduced by the refraction and is modeled using a 2-D Gaussian profile with $\sigma=10^\circ$
around each peak direction.  During the DOM deployment and freeze-in within the glacial ice sheet, the azimuthal orientations of the DOMs are not controlled and are initially unknown.
The orientation of each DOM, and therefore the initial direction of emitted light
from each LED, is determined to a precision of about $10^\circ$ by flashing individual horizontal LEDs
and studying the light arrival time at the six surrounding strings.  Here one relies on direct light from
an LED facing a target arriving sooner than scattered light from one facing away.

\section{Flasher data set}
\label{flasherdata}

The data set used in this paper includes at least 250 flashes from each DOM on string 63.  DOMs
were flashed at 1.2~Hz in a sequence, using a 70~ns pulse width and maximum brightness.  The six
horizontal LEDs on each flasher board were operated simultaneously, creating a pattern of light with approximate
azimuthal symmetry around the flasher string.  Flash sequences for DOMs at different depths were overlapping
but were sufficiently displaced in time that pulses of observed light were unambiguously assigned to individual flashers.

\begin{figure}[!h]\begin{center}
\begin{tabular}{c}
\mbox{\epsfig{file=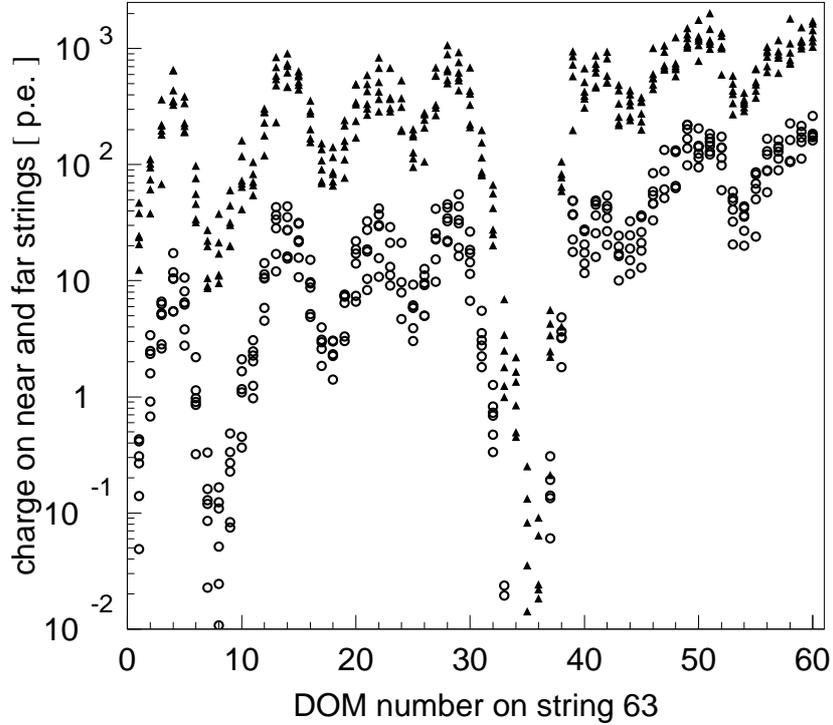,width=.8\textwidth}} \\
\end{tabular}
\parbox{.9\textwidth}{\caption{\label{mean}Charge collected by DOMs on the six nearest strings ($121.8-126.6$~m away, triangles) and six next-to-nearest strings ($211.4-217.9$~m away, circles), observed when flashing at the same position on string 63. }}
\end{center}\end{figure}

As seen in Fig.\ \ref{mean}, there is a substantial variation among the charges collected in DOMs at approximately the same depth as the emitter on the six surrounding strings. Some of the variation is due to relative differences in light yield between LEDs, and some is due to differences in distance to, and depth of, the six surrounding strings. Other reasons may include non-homogeneity of the ice.

The pulses corresponding to the arriving photons were extracted from the digitized waveforms and binned in 25~ns bins, from 0 to 5000~ns from the start of the flasher pulse (extracted from the special-purpose ATWD channel of the flashing DOM). To reduce the contribution from saturated DOMs (most of which were near the flashing DOM on string 63) \cite{pmt}, and to minimize the effects of the systematic uncertainty in the simulated angular sensitivity model of a DOM, the photon data collected on string 63 were not used in the fit. A DOM becomes saturated when it is hit by so many photons that the charge in its digitized output is no longer proportional to the number of incident photons.

The angular sensitivity model specifies a fraction of photons that are detected at a given angle with respect to the PMT axis. It accounts for the nominal DOM sensitivity measured in the lab, modified by the scattering in the column of re-frozen ice (see Fig.\ \ref{oms} and further discussion in section \ref{simulation}). Variations in the angular sensitivity model have a large impact on the simulated DOM response to the photons arriving along the PMT axis (straight into the PMT or into the back of a DOM), while the response to photons arriving from the sides of the PMT is much less affected.

\section{Six-parameter ice model}
\label{six}
This section overviews the ice parameterization introduced in \cite{kurt}, which in this paper is referred to as the six-parameter ice model. The ice is described by a table of depth-dependent parameters $b_e(400)$ and $a_{\rm dust}(400)$ related to scattering and absorption at a wavelength of 400~nm, by the depth-dependent relative temperature $\delta\tau$, and by the six global parameters (measured in \cite{kurt}): $\alpha$, $\kappa$, A, B, D, and E, which are described below. The thickness of the ice layers was somewhat arbitrarily chosen to be 10~m.
The scattering and absorption coefficients of each ice layer are best interpreted as the average of their true values over the thickness of the ice layer.
The chosen thickness of 10~m is the same as the value chosen in \cite{kurt} but smaller than the vertical DOM spacing of 17~m. Due to small depth offsets between the DOMs on different strings, we retain at least 1 receiving DOM per layer.

The geometrical scattering coefficient $b$ determines the average distance between successive scatters (as $1/b$). It is often more convenient to quote the effective scattering coefficient, $b_e=b\cdot(1-\langle\cos\theta\rangle)$, where $\theta$ is the deflection angle at each scatter. The absorption coefficient $a$ determines the average distance traveled by a photon before it is absorbed (as $1/a$).

The wavelength dependence of the scattering and absorption coefficients is given by the following expressions (for wavelength $\lambda$ in nm).
The power law dependence is predicted by theoretical models of light scattering in dusty ice. The power law dependence on photon wavelength was verified in the AMANDA study, using light sources with several different frequencies \cite{kurt}.
The effective scattering coefficient, with the global fit parameter $\alpha$, is
\[b_e(\lambda)=b_e(400)\cdot\left({\lambda\over 400}\right)^{-\alpha}.\]
The total absorption coefficient is the sum of two components, one due to dust and the other a temperature dependent component for pure ice \cite{kurt}:
\[a(\lambda)=a_{\rm dust}(\lambda)+Ae^{-B/\lambda}\cdot(1+0.01\cdot\delta\tau), \quad \mbox{with} \quad a_{\rm dust}(\lambda)=a_{\rm dust}(400)\cdot\left(\lambda\over 400\right)^{-\kappa}.\]

The parameter $\delta \tau$ above is the temperature difference relative to the depth of 1730~m (center of AMANDA):
\[\delta\tau(d)=T(d)-T(1730\ {\sf m}).\]
The temperature $T$(K) vs.\ depth $d$(m) is parameterized in \cite{temp} as:
\[T=221.5 - 0.00045319\cdot d + 5.822 \cdot 10^{-6} \cdot d^2.\]

The two remaining global parameters, $D$ and $E$, were defined in \cite{kurt} in a relationship establishing a correlation between absorption and scattering: $a_{\rm dust}(400)\cdot400^{\kappa}\approx D\cdot b_e(400)+E$, but are not used in this paper.

This work presents the measurement of the values of $b_e(400)$ and $a(400)$ based on data taken at a wavelength of 400~nm and relies on the six-parameter ice model described above to extrapolate scattering and absorption for wavelengths other than 400~nm.

\section{Simulation}
\label{simulation}
The detector response to flashing each of the 60 DOMs on string 63 generated a large data set that required very fast simulations such that many different sets of the coefficients $b_e(400)$ and $a_{\rm dust}(400)$ could be compared efficiently with the data.
A program called PPC (photon propagation code, see \apname\ref{ppc}), was written for this purpose. PPC propagates photons through ice described by a selected set of parameter values for $b_e(400)$ and $a_{\rm dust}(400)$ until they reach a DOM or are absorbed. When using PPC, no special weighting scheme was employed except that the spherical DOMs were scaled up in radius by a factor of 5 to 16, depending on the required timing precision\footnote{
Special care was taken to minimize any bias on photon arrival times by oversizing DOMs. First, we oversize DOMs in the direction perpendicular to the photon direction in order to avoid an artificially reduced propagation path before reaching the receiver. Still, in the worst case, an increase in size by a factor of 16 above to the nominal DOM dimensions may introduce a maximum bias of $(16-1)\cdot 16.51$~cm / 22~cm/ns=11.3~ns towards earlier arrival times (for a DOM with radius 16.51~cm and for speed of light in ice of 22~cm/ns). However, on average this error is smaller. An additional consideration is the overestimated loss of photons if they would get absorbed when entering an oversized DOM. Therefore we allow the photons to continue propagating even after hitting an oversized DOM.
}, and the number of emitted photons was scaled down by a factor of $5^2$ to $16^2$, corresponding to the increased area of the DOM.

\begin{figure}[!h]\begin{center}
\mbox{\epsfig{file=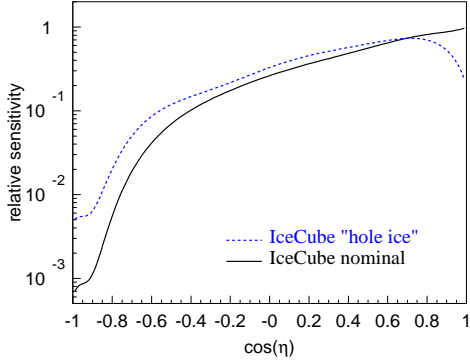,width=.45\textwidth}}
\parbox{.9\textwidth}{\caption{\label{oms} Angular sensitivity of an IceCube DOM where $\eta$ is the photon arrival angle with respect to the PMT axis. The nominal model, based on a lab measurement, is normalized to 1.0 at $\cos\eta=1$. The area under both curves is the same. }}

\end{center}\end{figure}
\begin{figure}[!h]\begin{center}
\begin{tabular}{ccc}
\mbox{\epsfig{file=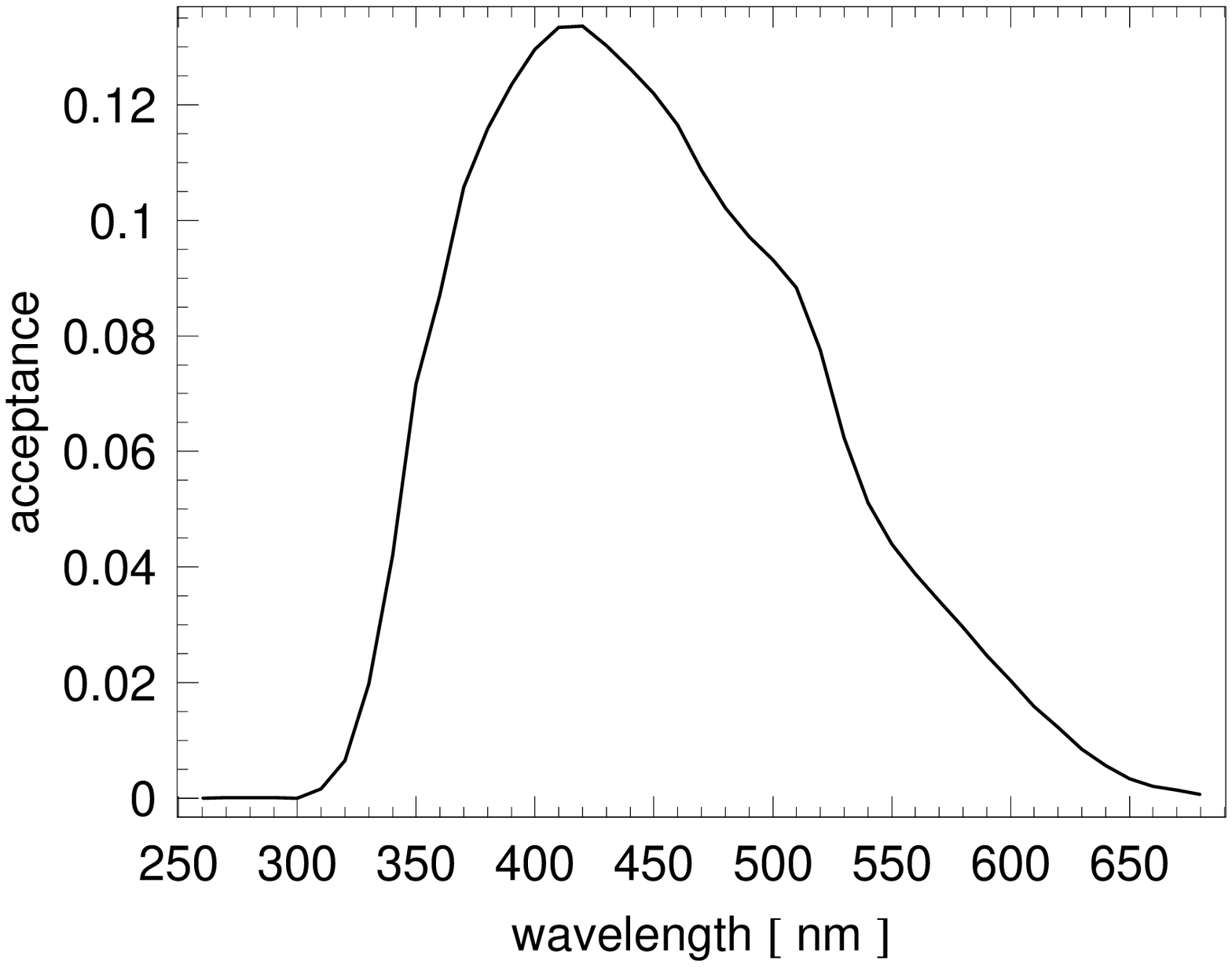,width=.45\textwidth}} & \ & \mbox{\epsfig{file=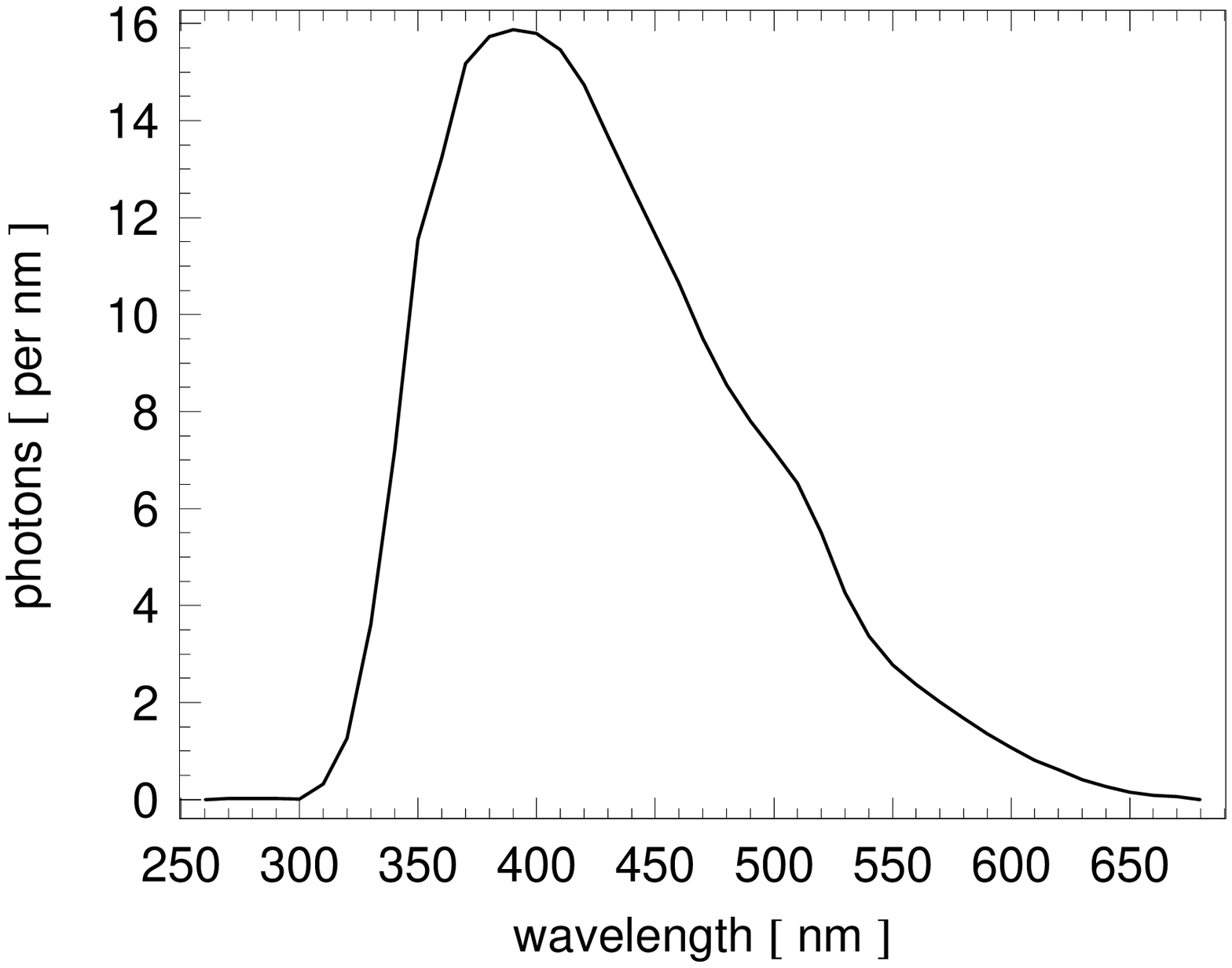,width=.45\textwidth}} \\
\end{tabular}
\parbox{.9\textwidth}{\caption{\label{oms2} Left: optical module acceptance: fraction of photons arriving from a direction parallel to the PMT axis (at $\cos\eta=1$) that are recorded. Note that the acceptance here is meant to include the glass and gel transmission and the PMT quantum and collection efficiencies. The acceptance is substantially lower at the peak than the roughly 20-25\% quantum efficiency of the PMT alone because it is given with respect to the photons incident on a cross-section of a DOM, which is larger than that of a PMT. Right: number of Cherenkov photons emitted by one meter of the simulated {\it bare} muon track (i.e., muon without secondary cascades), convolved with the optical module acceptance. The integral under the curve is 2450 photons. }}
\end{center}\end{figure}

The relative angular sensitivity of the IceCube DOM was modeled according to the ``hole ice'' description of \cite{effh}, which is shown in Fig.\ \ref{oms}. The ``hole ice,'' a column of ice approximately 30 cm in radius immediately surrounding the IceCube string, is described by taking into account an increased amount of scattering (with effective scattering length of 50~cm) via an empirical modification to the effective angular sensitivity curve of the receiving DOM.

The DOM acceptance is defined as the fraction of photons incident onto the cross-section of a DOM that cause a signal in the PMT. This fraction accounts for the losses due to the glass and gel transmission, and includes PMT quantum and collection efficiencies. It was calculated according to \cite{pmt} for a DOM of radius 16.51~cm. At 400~nm the DOM acceptance for the photons arriving at the PMT along its axis is 13.15\%. This corresponds to the nominal angular sensitivity curve of Fig.\ \ref{oms} peaking at 1.0 for $\cos\eta=1$. Additional considerations, including partial shadowing of the DOM surface by the supporting cables, lower this value by 10\%.

The actual number of isolated photoelectrons recorded by a DOM is reduced a further 15\% because of losses due to the discriminator threshold. The counting efficiency for single photons incident on a DOM is thus 13.15\% $\cdot$ 0.9 $\cdot$ (1-0.15) = 10\%. The peak of the amplitude distribution for one photoelectron is used to normalize this distribution and is henceforth used as a unit called p.e. The discriminator threshold is set at 0.25 p.e. The mean value of the amplitude distribution is found at 0.85 p.e., and at $N \cdot 0.85$ p.e. for $N$ photo-electrons recorded in one sensor. Thus, the fraction of charge recorded in multi-photoelectron records is the same as the recorded fraction of the number of isolated photoelectrons, 0.85. In a multi-photoelectron dominated situation this number can be used to convert from photoelectrons to amplitude in the p.e. unit. The product of this value and the two factors listed in the previous paragraph, 13.15\% $\cdot$ 0.9 $\cdot$ 0.85 = 10\%, is the ``effective acceptance,'' and is applied later (see section \ref{fitting}).

Naturally abundant cosmic ray muons which reach the depth of the detector produce Cherenkov light in a broad wavelength spectrum and may be used to test the ice model. For the tests presented in section \ref{comparison}, we simulate the light emitted by muons according to the following method. The Cherenkov photons were sampled from a convolution of the wavelength dependence of the DOM acceptance with the Cherenkov photon spectrum (see Fig.\ \ref{oms2} right) given by the Frank-Tamm formula \cite{tamm}:
\[{dN\over d\lambda dl}={2\pi\alpha\over \lambda^2}\sin^2\theta_c.\]
The muon light production is treated via the use of the ``effective length'' ($dl$), as described in \apname\ref{light}. The phase refractive index, $n_p$, used in the formula above (defining the Cherenkov angle $\cos\theta_c=1/n_p$) and the group refractive index, $n_g$, used in calculation of the speed of light in the medium, were estimated according to formulas from \cite{rind}:
\[n_p=1.55749-1.57988\cdot \lambda+3.99993\cdot \lambda^2-4.68271\cdot \lambda^3+2.09354\cdot \lambda^4\]
\[n_g=n_p\cdot (1+0.227106-0.954648\cdot \lambda+1.42568\cdot \lambda^2-0.711832\cdot \lambda^3).\]

\begin{figure}[!h]\begin{center}
\begin{tabular}{ccc}
\mbox{\epsfig{file=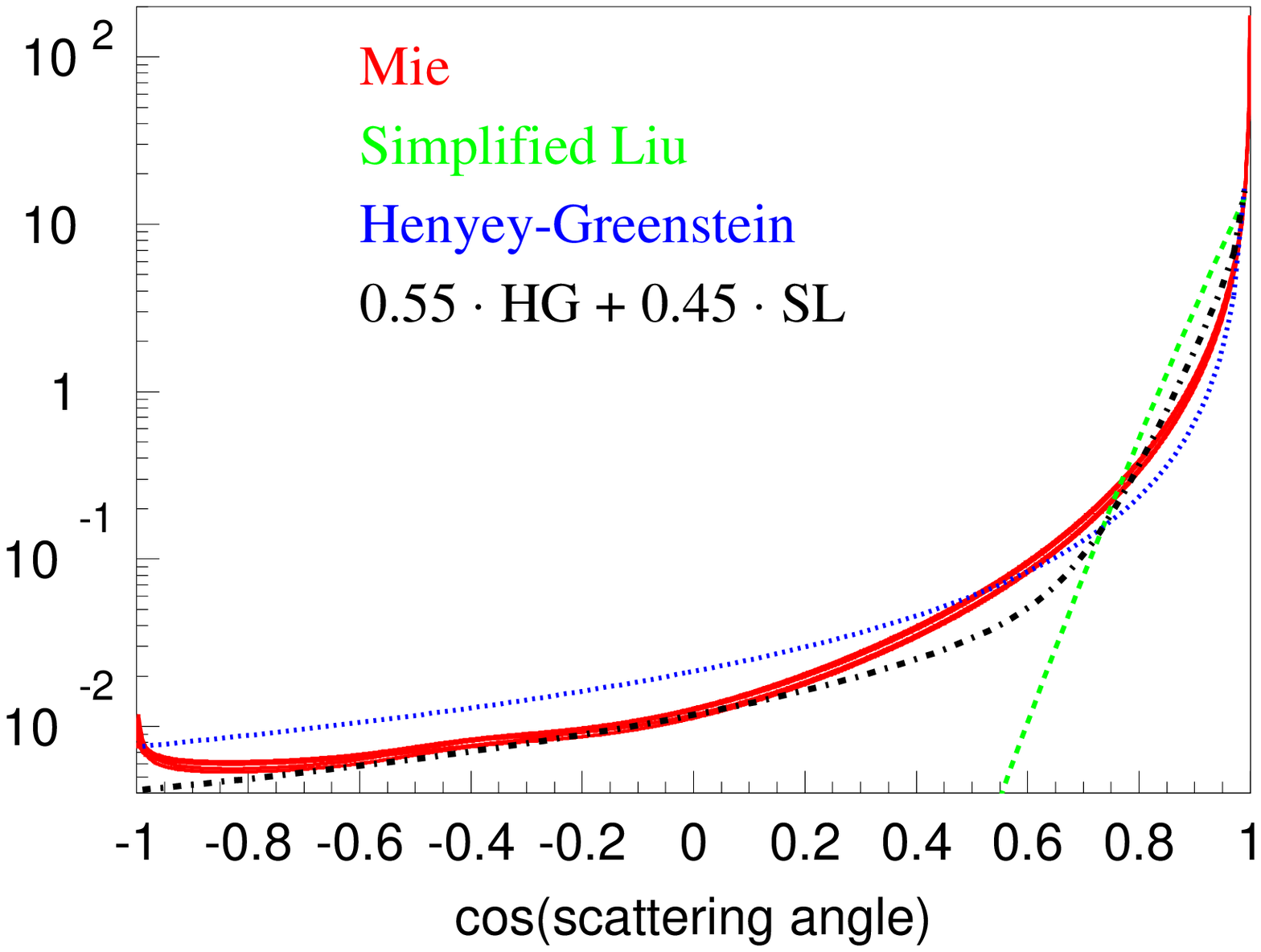,width=.45\textwidth}} & \ & \mbox{\epsfig{file=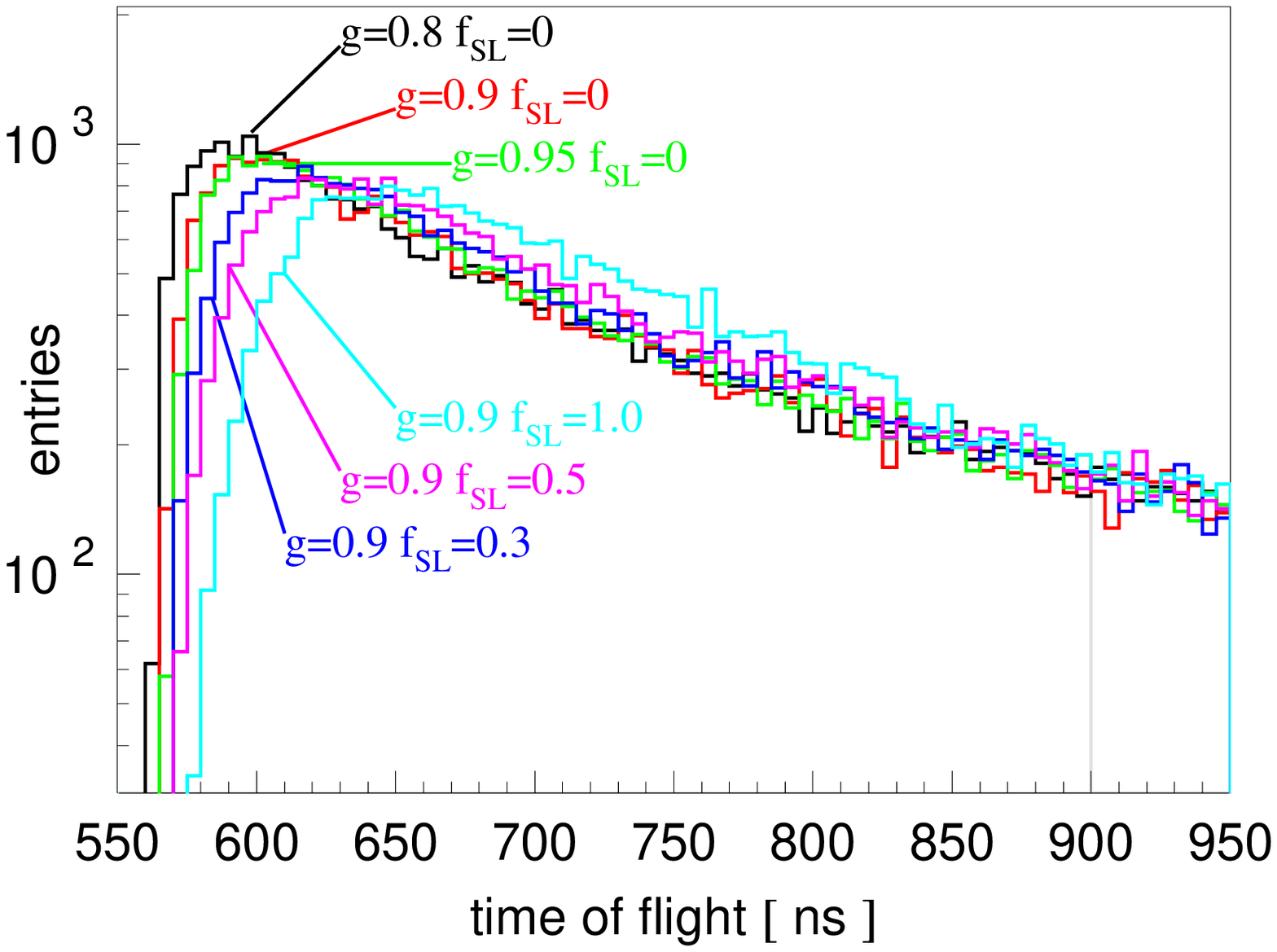,width=.45\textwidth}} \\
\end{tabular}
\parbox{.9\textwidth}{\caption{\label{mie} (left) Comparison of the Mie scattering profiles calculated at several depths of the South Pole ice with the Henyey-Greenstein (HG) \cite{kurt} and simplified Liu (SL) \cite{sam} scattering functions.  In each, $g=0.943$. }}
\parbox{.9\textwidth}{\caption{\label{shape} (right) Photon arrival time distributions at a DOM located 125~m away from the flasher, simulated for several values of $g=\langle\cos\theta\rangle$ and $f_{\sf SL}$. The difference in peak position simulated with $g=0.8$ and $g=0.9$ is of the same order ($\sim 10$ ns) as that between sets simulated with different values of the shape parameter $f_{\sf SL}$. }}
\end{center}\end{figure}

The distribution of the photon scattering angle $\theta$ is modeled by a linear combination of two functions commonly used to approximate scattering on impurities:
\[p(\cos\theta)=(1-f_{\sf SL})\cdot{\sf HG}(\cos\theta)+f_{\sf SL}\cdot {\sf SL}(\cos\theta).\]
The first is the Henyey-Greenstein (HG) function \cite{kurt}:
\[{\sf HG}(\cos\theta)={1\over 2}{1-g^2\over [1+g^2-2g\cdot\cos\theta]^{3/2}}, \quad {\sf with} \quad g=\langle\cos\theta\rangle,\]
which can be analytically integrated and inverted to yield a value of $\cos\theta$ as a function of a random number $\xi$ uniformly distributed on interval $[0; 1]$:
\[\cos\theta={1\over 2g}\left(1+g^2-\left({1-g^2\over 1+gs}\right)^2\right), \quad s=2\cdot \xi-1.\]
The second is the simplified Liu (SL) scattering function \cite{sam}:
\[{\sf SL}(\cos\theta) = {1+\alpha\over 2} \cdot \left({1+\cos\theta\over 2}\right)^\alpha, \quad \mbox{with} \quad \alpha={2g\over 1-g}, \quad g=\langle\cos\theta\rangle,\]
which also yields a simple expression for $\cos\theta$ as a function of a random number $\xi\in [0; 1]$:
\[\cos\theta=2\cdot\xi^\beta-1, \quad \mbox{with} \quad \beta={1-g\over 1+g}.\]
Figure \ref{mie} compares these two functions with the prediction of the Mie theory, with dust concentrations and radii distributions taken as described in \cite{kurt}. The photon arrival time distributions are substantially affected by the ``shape'' parameter $f_{\sf SL}$ (as shown in Fig.\ \ref{shape}), making it possible to determine this parameter from fits to data.

A value of $g=0.9$ was used in this work (cf.\ $g=0.8$ in \cite{kurt}).
A higher value, 0.94, is predicted by Mie scattering theory \cite{kurt,mie},
%Higher values, to a maximum of approximately 0.94 \cite{kurt,mie}, are predicted by the Mie scattering theory
but results in slower simulation and almost unchanged values of the effective scattering ($b_e$) and absorption ($a$) coefficients, as shown in \cite{kurt}.

\section{Likelihood description}
\label{llh}
Consider the amount of charge received by DOM $i$ in time bin $n$ when flashing DOM $k$. This charge is measured by taking data with a total photon count of $d$ in $n_d$ flasher events and a per-event expectation of $\mu_d$. This charge is predicted by the simulation with a total photon count of $s$ in $n_s$ simulated events and a per-event expectation of $\mu_s$. Naively one expects the best approximations to $\mu_d$ and $\mu_s$ from data and simulated events to be $\mu_d=d/n_d$ and $\mu_s=s/n_s$.

The error in describing data with simulation (i.e., describing $\mu_d$ with $\mu_s$) is approximately $20-30\%$ (estimated as described later in section \ref{errors}). One quantifies the amount of disagreement between data and simulation in the presence of such an error with a $\chi^2_{i,n,k}$. Omitting the indices $i$, $n$, and $k$, this is given by:
\[\chi^2={(\ln\mu_d - \ln\mu_s)^2 \over \sigma^2 }.\]
The uncertainty due to this systematic error can be modeled with a probability distribution function
\[{1 \over \sqrt{2\pi} \sigma} \exp{-(\ln\mu_d - \ln\mu_s)^2 \over 2 \sigma^2}.\]

Given that $\mu_d$ and $\mu_s$ are not known, and the measured values are $d$ and $s$, one formulates the likelihood function that describes counts measured in both data and simulation as
\[{(\mu_sn_s)^s \over s!} e^{-\mu_sn_s} \cdot {(\mu_dn_d)^d \over d!} e^{-\mu_dn_d} \cdot {1 \over \sqrt{2\pi} \sigma} \exp{-(\ln\mu_d - \ln\mu_s)^2 \over 2 \sigma^2}.\]
Taking the negative logarithm, this becomes:
\[\ln s! + \mu_sn_s -s\ln(\mu_sn_s) + \ln d! + \mu_dn_d -d\ln(\mu_dn_d) + {1 \over 2\sigma^2} \ln^2 {\mu_d \over \mu_s} +\ln(\sqrt{2\pi} \sigma)\equiv F.\]

The function $F(\mu_s, \mu_d)$ can be easily minimized against $\mu_s$ and $\mu_d$, yielding estimates of these quantities. To demonstrate this, the derivatives of $F$ are first calculated and set to 0:
\[\mu_s {\partial F \over \partial \mu_s} = \mu_sn_s - s - {1 \over \sigma^2} \ln{\mu_d \over \mu_s} = 0,\]
\[\mu_d {\partial F \over \partial \mu_d} = \mu_dn_d - d + {1 \over \sigma^2} \ln{\mu_d \over \mu_s} = 0.\]
The sum of these ($\mu_sn_s + \mu_dn_d = s+d$) yields an expression of $\mu_d$ as a function of $\mu_s$. Plugging it into the first of the above two equations one gets
\[f=\mu_s {\partial F \over \partial \mu_s} (\mu_s, \mu_d(\mu_s)) = \mu_sn_s - s - {1 \over \sigma^2} \ln{\mu_d(\mu_s) \over \mu_s} = 0.\]

This equation can be solved with a few iterations of the Newton's root finding method starting with a solution to
\[\mu_s=\mu_d(\mu_s)\mbox{:} \quad \quad \mu_s = \mu_d = {s+d \over n_s+n_d}.\]
At each iteration the value of $\mu_s$ is adjusted by $-f/f^\prime$, where the derivative is evaluated as
\[f^\prime=n_s\left( 1+{1 \over \sigma^2} ({1 \over \mu_sn_s} + {1 \over \mu_dn_d})\right).\]

Once the likelihood function is solved for the best values of $\mu_s$ and $\mu_d$, they may be inserted into the expression for $\chi^2_{i,n,k}$ above. One can now write the complete $\chi^2$ function (adding the regularization terms $R_j$ described in the next section) as a sum over DOMs $i$, used in the analysis, multiplied by time bins $n$, when flashing DOMs $k$:
\[\chi^2=\sum_{i,n,k}{(\ln\mu_d - \ln\mu_s)^2 \over \sigma^2 } + \sum_{j=\{r,u\}}\alpha_jR_j.\]

\section{Regularization terms}
\label{reg}
Two regularization terms (see, e.g., \cite{svd}) are added to the likelihood function described in the previous section. The first term suppresses the fluctuations of scattering and absorption coefficients with depth in under-constrained ice layers. It is formed from terms that are numerical expressions for second derivatives of scattering and absorption with respect to the position of the ice layer:
\bal
R_r=\sum_{i=2}^{N-1} & \left[ (\ln b_e[i-1]-2\cdot\ln b_e[i]+\ln b_e[i+1])^2 \right. \\
& \left. + (\ln a_{\rm dust}[i-1]-2\cdot\ln a_{\rm dust}[i]+\ln a_{\rm dust}[i+1])^2\right].
\eal
Here $N$ is the number of ice layers in which $b_e$ and $a_{\rm dust}$ are defined.

The second term is used to suppress fluctuations in the diagram of $a_{\rm dust}$ vs.\ $ b_e$, enforcing the notion that both are proportional to the dust concentration. It is constructed as an excess of the sum of distances between the consecutive points $(\ln b_e, \ln a_{\rm dust})$ over the shortest distance connecting the end points:
\[R_u=-D(1, N)+\sum_{j=1}^{N-1} D(j, j+1),\]
\[\mbox{where}\quad D(j_1, j_2)=\sqrt{(\ln b_e[j_1]-\ln b_e[j_2])^2 + (\ln a_{\rm dust}[j_1]-\ln a_{\rm dust}[j_2])^2}.\]
The points $(\ln b_e, \ln a_{\rm dust})$ are sorted by the value of $\ln b_e + \ln a_{\rm dust}$ and shown in the above sum with the index $j[i]$.

Both terms affect the resulting scattering and absorption coefficients by on average less than 2\% at detector depths at their chosen strengths $\alpha_{r,u}$. Deviations larger than this, up to 19\% were observed in the region of particularly dusty ice around the depth of 2000~m. The size of the effect has been verified by re-running the fits without including the terms. The regularization terms are likely to become more important if the thickness of ice layers (10~m in this work) were chosen to be much smaller than the spacing between DOMs on a string (17 m).

\section{Fitting the flasher data}
\label{fitting}
The six horizontal LEDs within a single DOM flashing at maximum brightness and width nominally emit about $4.5\cdot 10^{10}$ photons \cite{perf}. After accounting for the effective DOM acceptance (as explained in section \ref{simulation}), these photons result in a charge amplitude of $4.5\cdot 10^9$ p.e., which henceforth is traced as $4.5\cdot 10^9$ ``photons'' that each result in an amplitude of 1 p.e. Using a DOM size scaling factor of 16, only $1.76\cdot 10^7$ photons need to be simulated ($16^2=256$ times fewer).

Simulating 9765625 photons, with a scaling factor of 16, corresponds to $2.5\cdot 10^9$ photons simulated for actual-size receiving DOMs, or $2.5\cdot 10^{10}$ real photons leaving the flasher DOM (after accounting for the effective acceptance of the receiving DOM). This is defined as a ``unit bunch'' of photons, which is simulated in approximately 1 second on a single GPU (see \apname\ref{ppc}).

In the following discussion, a ``photon yield factor'' ($p_y$) is the number of unit bunches that corresponds to a given number of real photons. For instance, $4.5\cdot 10^{10}$ photons emitted by a flasher board correspond to a photon yield factor of $p_y=1.8$.

Data from all pairs of emitter-receiver DOMs (located in the same or different ice layers, amounting to about 38700 pairs) contributed to the fit to 200 ice parameters (scattering and absorption in 10~m layers at detector depths of 1450 to 2450 m). Two $\chi^2$ functions were used in fitting the data: $\chi^2_q$ was constructed with one term from each emitter-receiver pair (using the total recorded charge) and $\chi^2_t$ was constructed with the recorded charge split in 25~ns bins. Although $\chi^2_t$ more completely used the available information, $\chi^2_q$ was found to be somewhat more robust with respect to statistical fluctuations in the simulated sets and was faster to compute. Thus, $\chi^2_q$ was used in an initial search for a solution, with $\chi^2_t$ applied in the final fits.

Both $b_e(400)$ and $a_{\rm dust}(400)$ are roughly proportional to the concentration of dust (this would be precise if the dust composition in the ice were the same at all depths). This motivates the following simplification in the initial search for the minimum of $\chi^2_q$: in each layer both $b_e(400)$ and $a_{\rm dust}(400)$ are scaled up or down by the same relative amount, ranging from 1-40\%, preserving their ratio to each other.

\begin{table}[!h]\begin{center}
\begin{tabular}{|c|}
\hline
Starting with some initial values of $b_e(400) \propto a_{\rm dust}(400)$ and some $p_y$, $t_{\sf off}$, $f_{\sf SL}$: \\
\begin{tabular}{|l|}
\hline
{\it Using} $\chi^2_q$ find best values of $b_e(400) \sim a_{\rm dust}(400)$ \\
{\it Using} $\chi^2_t$ find best values of $p_y$, $t_{\sf off}$, $f_{\sf SL}$, $\alpha_{\rm sca}$, $\alpha_{\rm abs}$: \\
\begin{tabular}{c||cl}
 & $p_y$: & photon yield factor \\
 & $t_{\sf off}$: & global time offset for all flasher pulses \\
 & $f_{\sf SL}$: & shape parameter of the scattering function \\
 & $\alpha_{\rm sca}$: & scaling of scattering coefficient table \\
 & $\alpha_{\rm abs}$: & scaling of absorption coefficient table \\
\end{tabular} \\
repeat this box until converged ($\sim 3$ iterations) \\
\hline
\end{tabular} \\
{\it Using} $\chi^2_t$, refine the fit with $b_e(400)$ and $a_{\rm dust}(400)$ fully independent from each other. \\
\hline
\end{tabular}
\caption{\label{glob}Flow chart of the global fit procedure to ice/flasher parameters.}
\end{center}\end{table}

\begin{figure}[!h]\begin{center}
\begin{tabular}{ccc}
\mbox{\epsfig{file=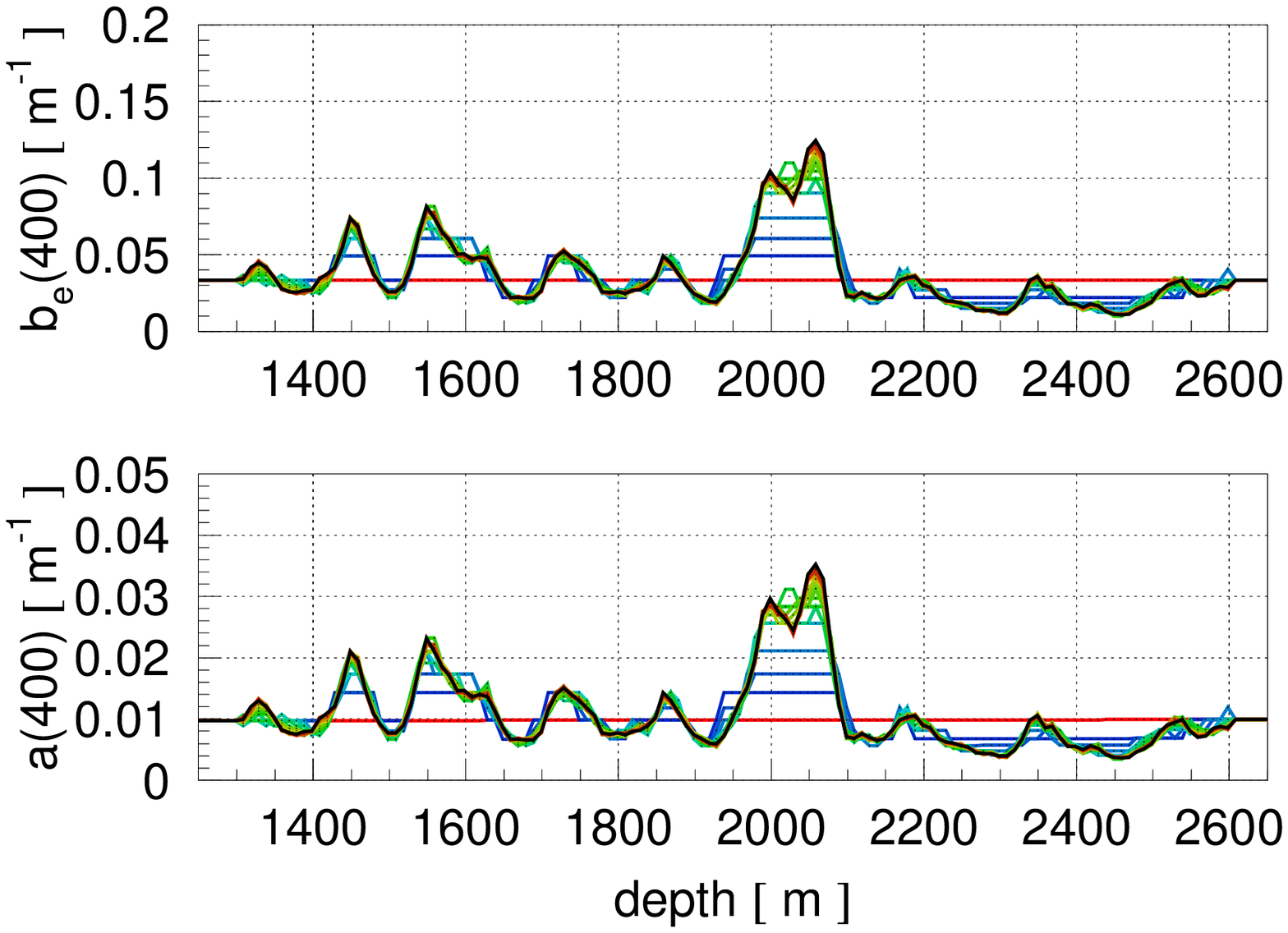,width=.45\textwidth}} & \ & \mbox{\epsfig{file=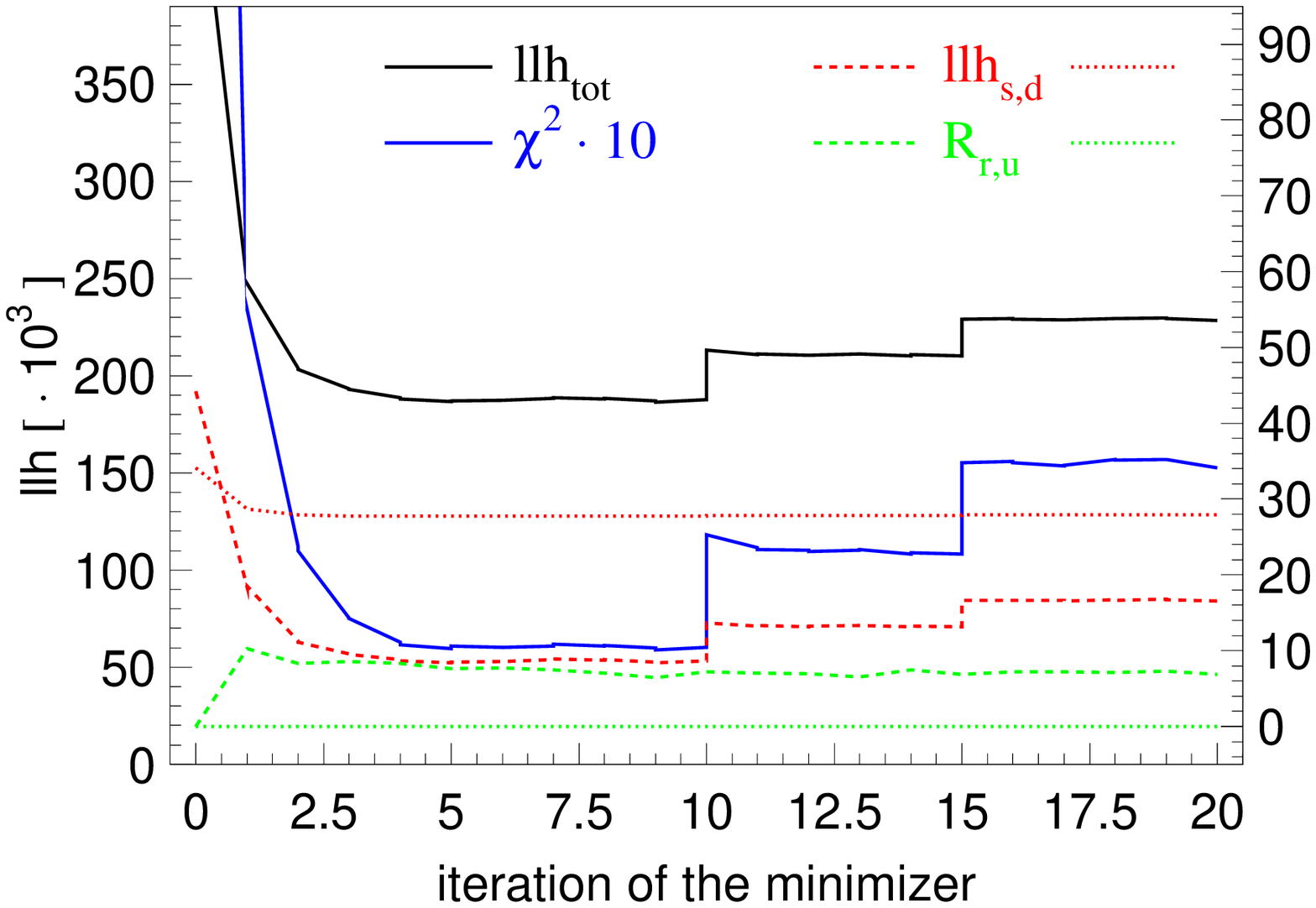,width=.45\textwidth}} \\
\end{tabular}
\parbox{.9\textwidth}{\caption{\label{conv} Left: values of $b_e(400)$ and $a(400)$ vs.\ depth after 20 steps of the minimizer. The black curve shows fitted values after the last step of the minimizer. Right: global $\chi^2_q$ values achieved after each step of the minimizer. The starting ``homogeneous ice'' value is $1.34\cdot 10^5$. Regularization terms $R_{r,u}$ use the scale on the right. Also shown are the Poisson terms for simulation and data (llh$_{\rm s,d}$) and the full likelihood including all terms (llh$_{\rm tot}$). The $\chi^2$ changes suddenly when the number of simulated flasher events is increased, but ultimately decreases as the minimizer steps through the iterations. Note that for iteration steps 1-10, only 1 flasher event is simulated. For steps 11-15 and steps 16-20, 4 events and 10 events, respectively, are simulated.}}
\end{center}\end{figure}

Starting with homogeneous ice described by $b_e(400)=0.042$ m$^{-1}$ and $a_{\rm dust}(400)=8.0$ km$^{-1}$ (average of \cite{kurt} at detector depths), the minimum of $\chi^2_q$ is found in about 20 steps. At each iteration, the values of $b_e(400)$ and $a_{\rm dust}(400)$ are varied across consecutive ice layers, one layer at a time. Five flashing DOMs closest to the layer in which the properties are varied are used to estimate the variation of the $\chi^2$. Figure \ref{conv} shows fitted ice properties after each of 20 steps of the minimizer.

\begin{figure}[!h]\begin{center}
\begin{tabular}{ccc}
\mbox{\epsfig{file=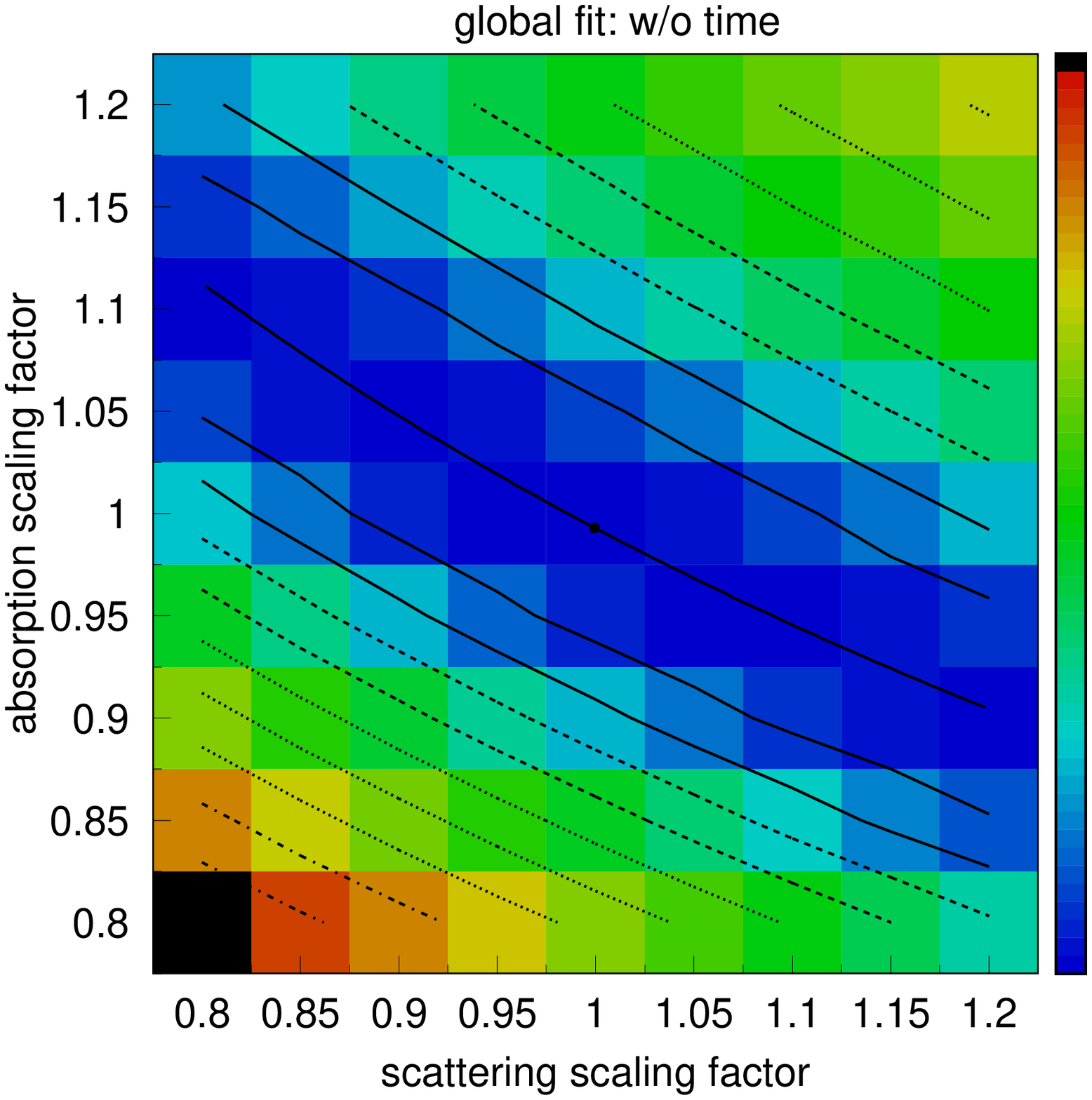,width=.45\textwidth}} & \ & \mbox{\epsfig{file=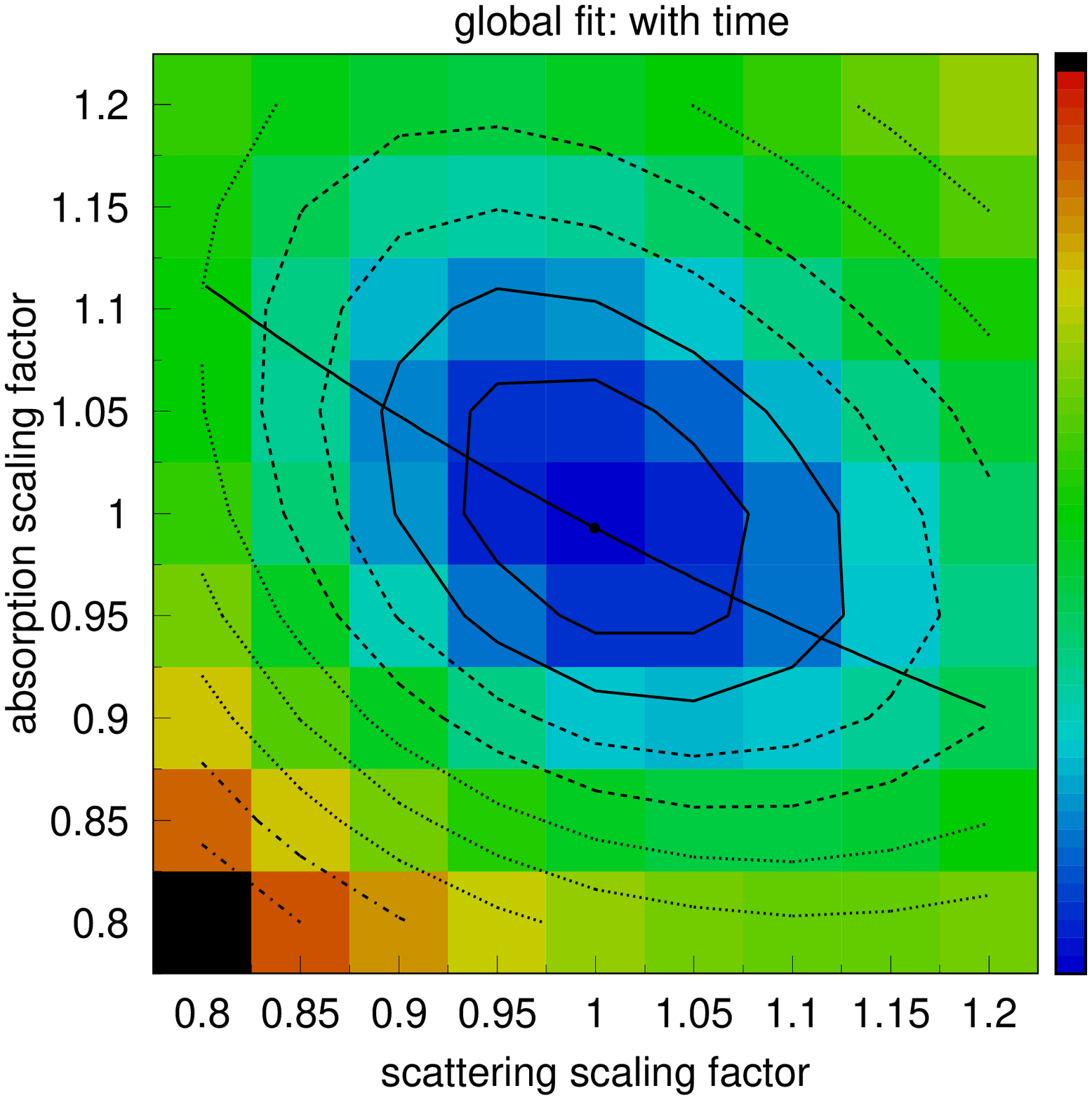,width=.45\textwidth}} \\
\end{tabular}
\parbox{.9\textwidth}{\caption{\label{min} Likelihood functions in the vicinity of their minima constructed using only charge information (left), and using full timing information (right). The values are shown on a log scale (with colors and contours). The ranges of values shown are: $\chi^2_q=1.43\cdot 10^4$ to $1.51\cdot 10^5$ (left) and $\chi^2_t=1.05\cdot 10^5$ to $4.01\cdot 10^5$ (right). }}
\end{center}\end{figure}

The search for the minimum of $\chi^2_t$ is performed next in the parameter space of the overall time offset from the flasher start time ($t_{\sf off}$), photon yield factor ($p_y$), shape parameter ($f_{\sf SL}$) of the scattering function (see section \ref{simulation}), and scaling coefficients $\alpha_{\rm sca}$ and $\alpha_{\rm abs}$ applied to the depth tables of $b_e(400)$ and $a_{\rm dust}(400)$.

The $b_e(400)$ and $a_{\rm dust}(400)$ of the solution are scaled with coefficients $\alpha_{\rm sca}$ and $\alpha_{\rm abs}$ to produce the likelihood profiles shown in Fig.\ \ref{min}. From this figure, it is apparent that using the timing information is necessary to resolve both $b_e(400)$ and $a_{\rm dust}(400)$. The minimum of $\chi^2_q$ has an elongated shape, and the direction of its longest extension is determined. The point on the line drawn in this direction through the minimum of $\chi^2_q$ is chosen to minimize the $\chi^2_t$. The global scaling factors $\alpha_{\rm sca}$ and $\alpha_{\rm abs}$ corresponding to this point are used to rescale the starting ``homogeneous ice'' values of $b_e(400)$ and $a_{\rm dust}(400)$. The entire procedure is then repeated.

The solution is finally refined by varying $b_e(400)$ and $a_{\rm dust}(400)$ at each step of the $\chi^2_t$ minimizer four times (combinations of $b_e\pm\delta b_e$ and $a_{\rm dust}\pm\delta a_{\rm dust}$, with $\delta b_e/b_e$ and $\delta a_{\rm dust}/a_{\rm dust}=1-2$\%). The entire procedure described above is also outlined in Table \ref{glob}.

\begin{figure}[!h]\begin{center}
\mbox{\epsfig{file=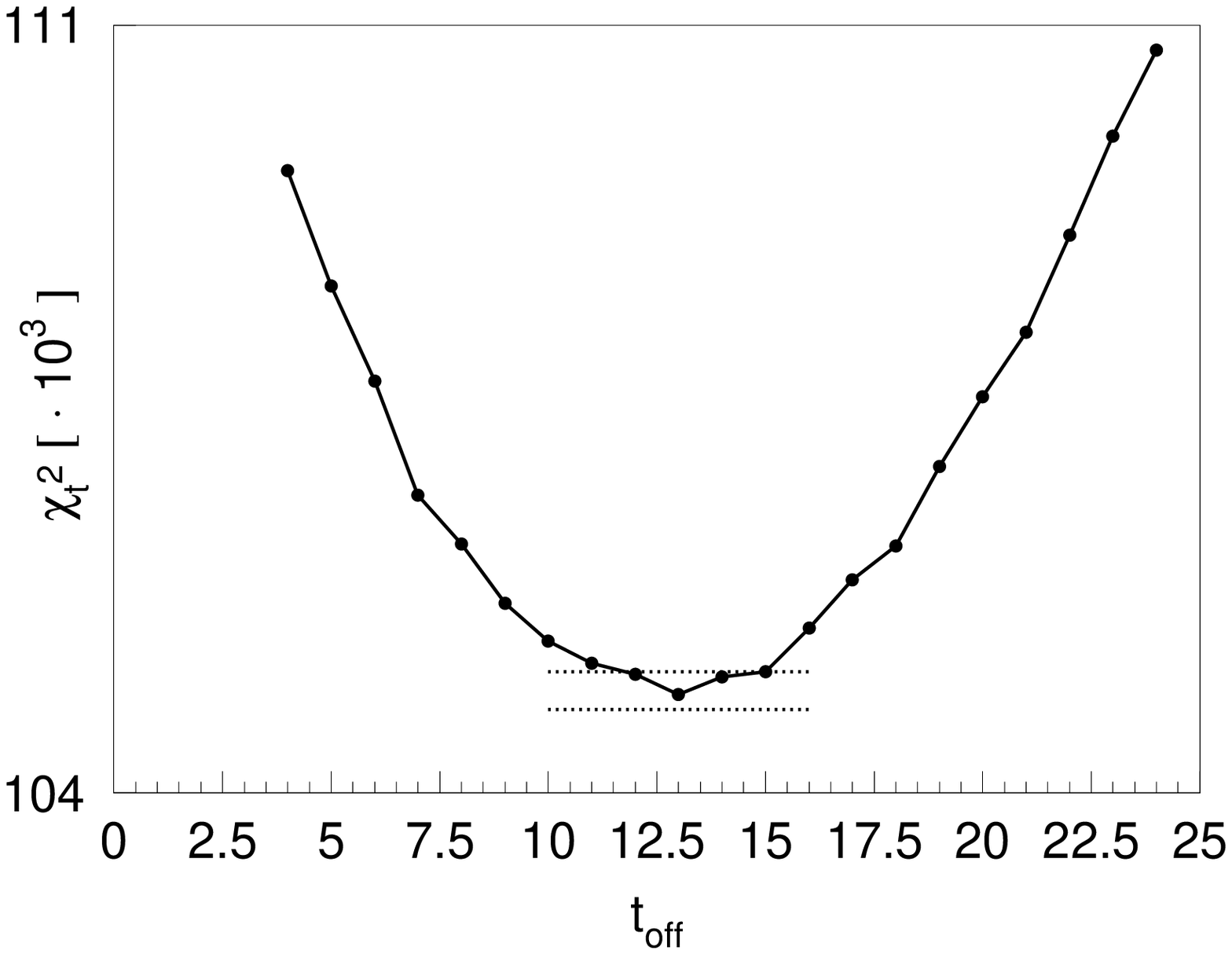,width=.45\textwidth}}
\begin{tabular}{ccc}
\mbox{\epsfig{file=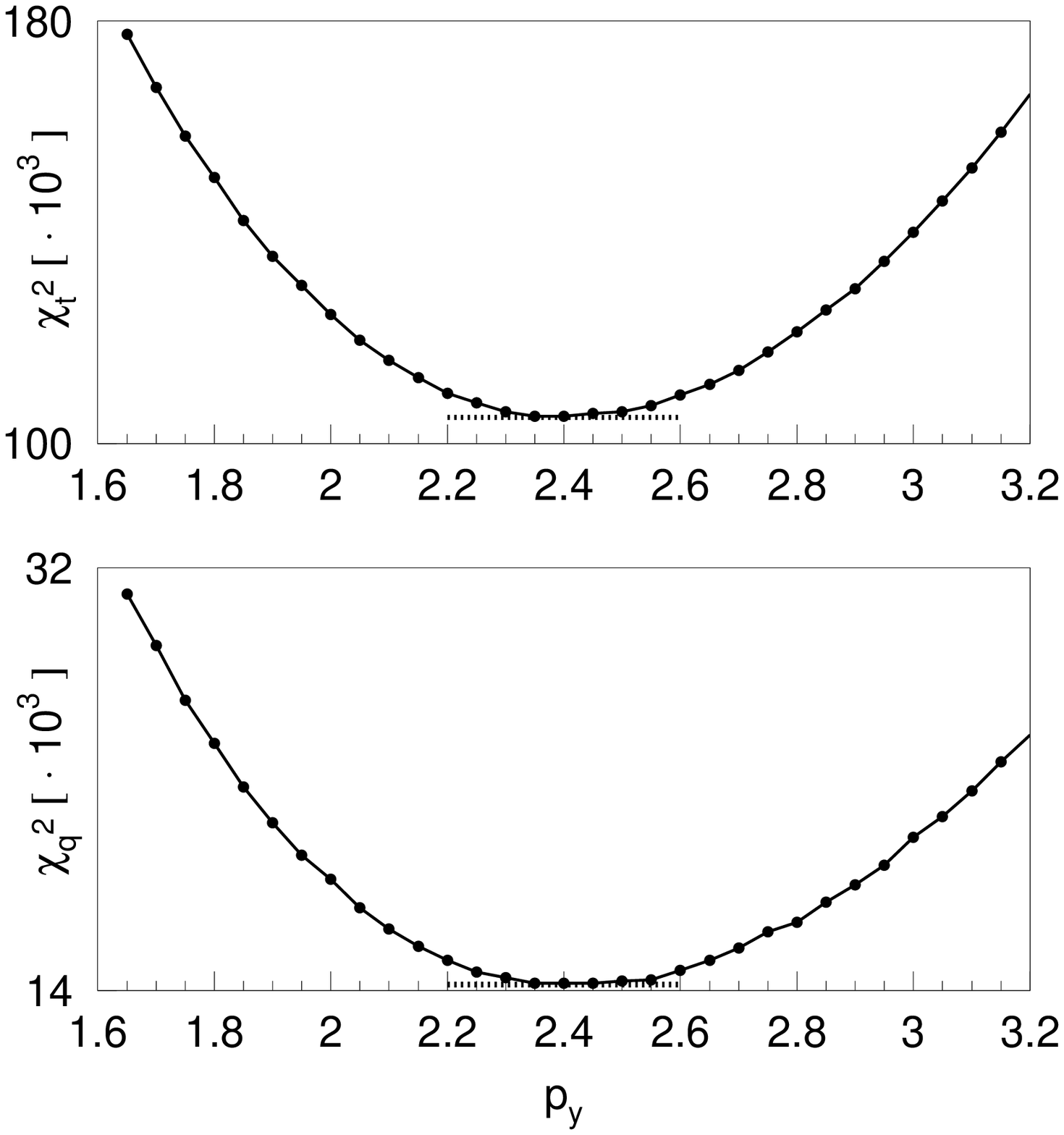,width=.45\textwidth}} & \ & \mbox{\epsfig{file=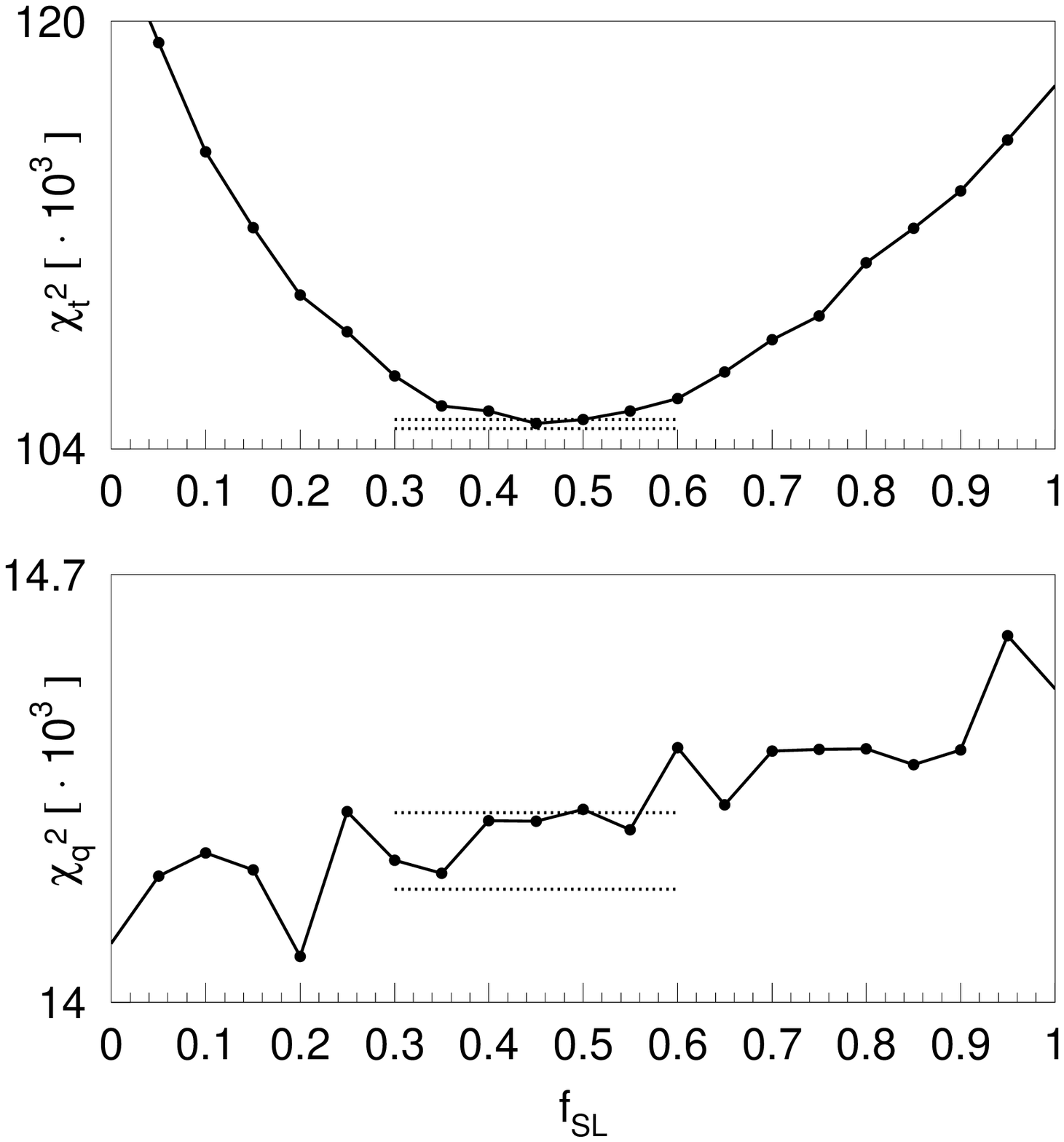,width=.45\textwidth}} \\
\end{tabular}
\parbox{.9\textwidth}{\caption{\label{py} Behavior of $\chi^2_t$ and $\chi^2_q$ in the vicinity of the fit minimum vs.\ $t_{\sf off}$, $p_y$, and $f_{\sf SL}$. All plots are shown on a linear scale. Horizontal dotted lines show the $\pm 1\sigma$ range due to purely statistical fluctuations in the simulation estimated for the best-fit model. The minimum in $t_{\sf off}$ and $f_{\sf SL}$ is a feature of $\chi^2_t$ but not $\chi^2_q$.}}
\end{center}\end{figure}

The best fit is achieved for $p_y=2.40\pm 0.07$.
Since the best value of $p_y$ is also calculated by the method above, the resulting table of $b_e(400)$ and $a_{\rm dust}(400)$ values is independent of a possible constant scaling factor in the charge estimate or the absolute sensitivity of a DOM. The best fit values of the other parameters are $t_{\sf off}=13\pm 2$ ns
and $f_{\sf SL}=0.45\pm 0.05$ (see Fig.\ \ref{py}). The typical agreement of data and simulation based on these parameters is demonstrated in Fig.\ \ref{schema}b.

\section{Dust logger data}
\label{logger}

Several dust loggers \cite{ryan} were used during the deployment of six of the IceCube strings resulting in a survey of the structure of ice dust layers with an effective resolution of approximately 2 mm. These layers were then matched across the detector to provide a {\it tilt map} of the South Pole ice, as well as a high-detail {\it average dust log} (a record of a quantity proportional to the dust concentration vs.\ depth). Additionally, the East Dronning Maud Land (EDML, see \cite{ryan}) ice core data were used to extend the dust record to below the lowest dust-logger-acquired point (taken at a depth of 2478 m).

\begin{figure}[!h]\begin{center}
\begin{tabular}{c}
\mbox{\epsfig{file=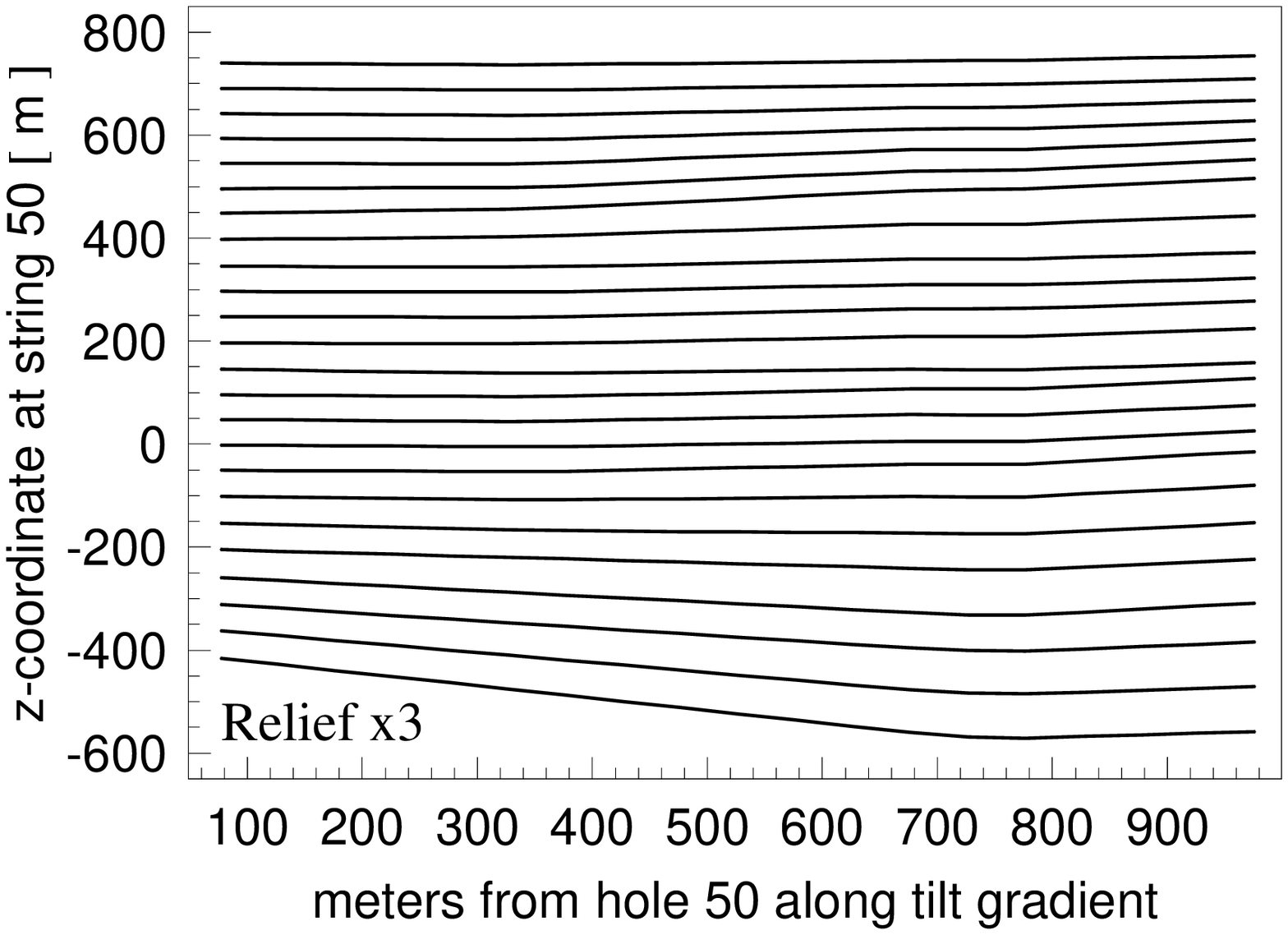,width=.65\textwidth}} \\
\mbox{\epsfig{file=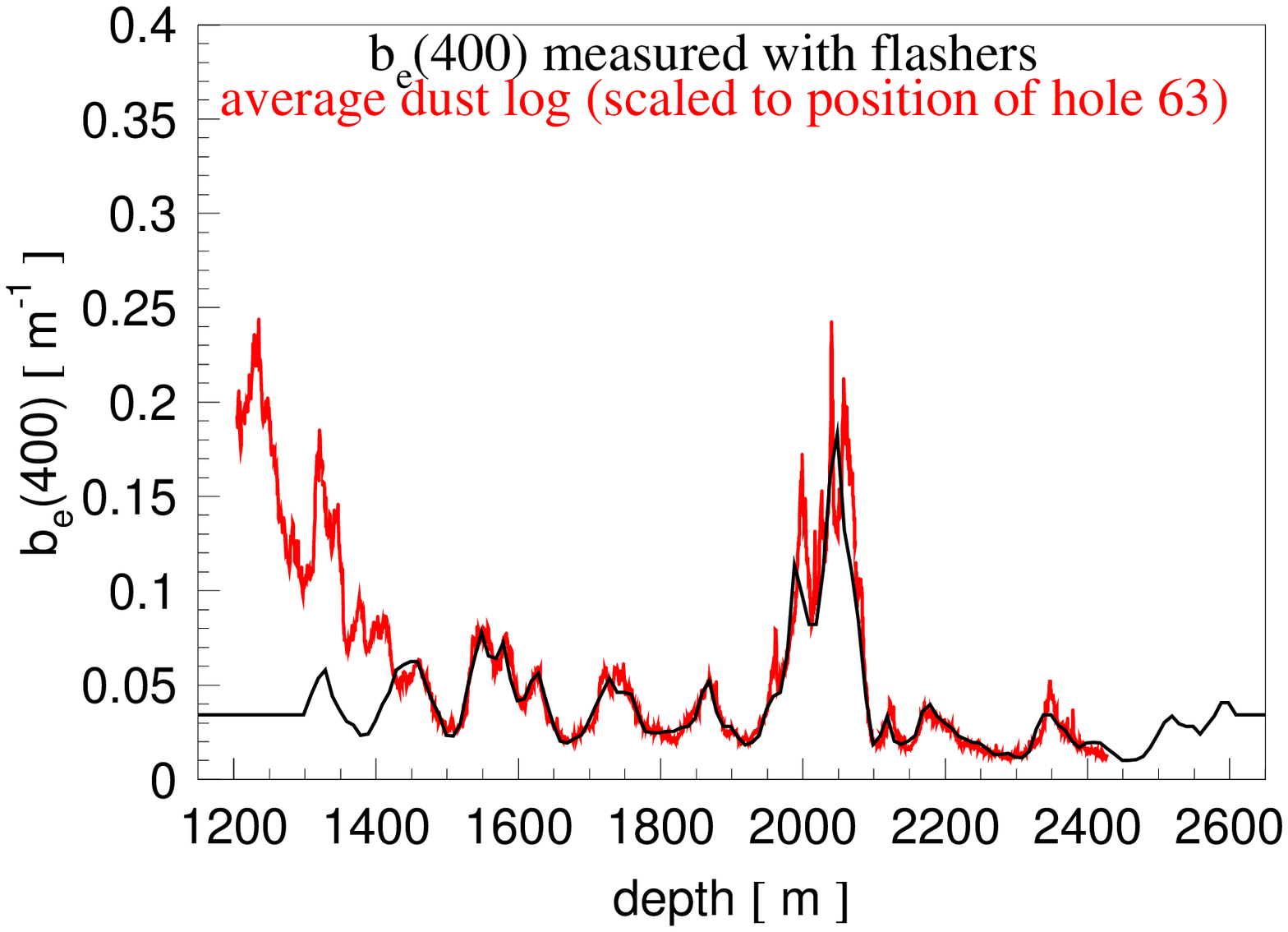,width=.65\textwidth}} \\
\end{tabular}
\parbox{.9\textwidth}{\caption{\label{tilt} Top: extension of ice layers along the average gradient direction. Relief is amplified by a factor of 3 to enhance the clarity of the layer structure. The lowest layer shown exhibits a shift of 56 meters between its shallowest and deepest points (which is the largest shift of all layers shown in the figure). Bottom: comparison of the average dust log with the effective scattering coefficient $b_e(400)$ measured with the flasher data. }}
\end{center}\end{figure}

The table of dust layer elevations (the {\it tilt map}) taken from \cite{ryan} provides the layer shift (relief) from its position at the location of a reference string, at a point distance $r$ away from this string, along the average gradient direction of 225 degrees SW (see Fig.\ \ref{tilt}). The $z$-coordinate of a given layer at $r$ is given by $z_r=z_0+{\rm relief}(z_0, r)$. Between the tabulated points, $z_r$ was calculated by linear interpolation in $z_0$ and $r$. The equation was solved by simple iteration resulting in a table of $z_0(z_r)-z_r$ vs.\ $z_r$ given at several points along the gradient direction. Combined with the dust depth record at the location of the reference string (at $r=0$), this yields a complete description of the dust profile in and around the detector (assuming that the concentration of dust is maintained along the layers).

The correlation between the effective scattering coefficient measured with the IceCube flasher data and the average dust log (scaled to the location of string 63) is excellent, as shown in Fig.\ \ref{tilt}. All major features agree, with well-matched rising and falling behavior, and are of the same magnitude. Some minor features are washed out in the flasher measurement.

With an established correlation to the average dust log, the EDML-extended version of the log was utilized in constructing an initial approximation (replacing the ``homogeneous ice'') used by the fitting algorithm described in section \ref{fitting}. This resulted in the recovery and enhancement of several features in the scattering and absorption vs.\ depth that were previously unresolved. Additionally, the solution is now biased towards the scaled values of the extended log (instead of the somewhat arbitrary values of the initial homogeneous ice approximation) in the regions where the flasher fitting method has no resolving power, i.e., above and below the detector.

\section{Uncertainties of the measurement and final result}
\label{errors}

\begin{figure}[!h]\begin{center}
\begin{tabular}{c}
\mbox{\epsfig{file=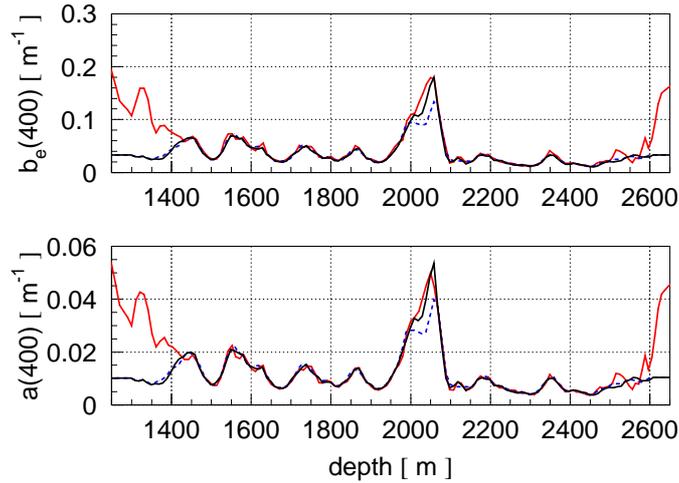,width=.65\textwidth}} \\
\end{tabular}
\parbox{.9\textwidth}{\caption{\label{sim} Reconstructed ice properties in black for simulated flasher events with input ice properties in red. The ice properties in the {\it dust peak} are reconstructed correctly with 250 simulated events per flasher. The blue dashed curve shows the result achieved with only 10 simulated events per flasher. }}
\end{center}\end{figure}

\begin{figure}[!h]\begin{center}
\begin{tabular}{c}
\mbox{\epsfig{file=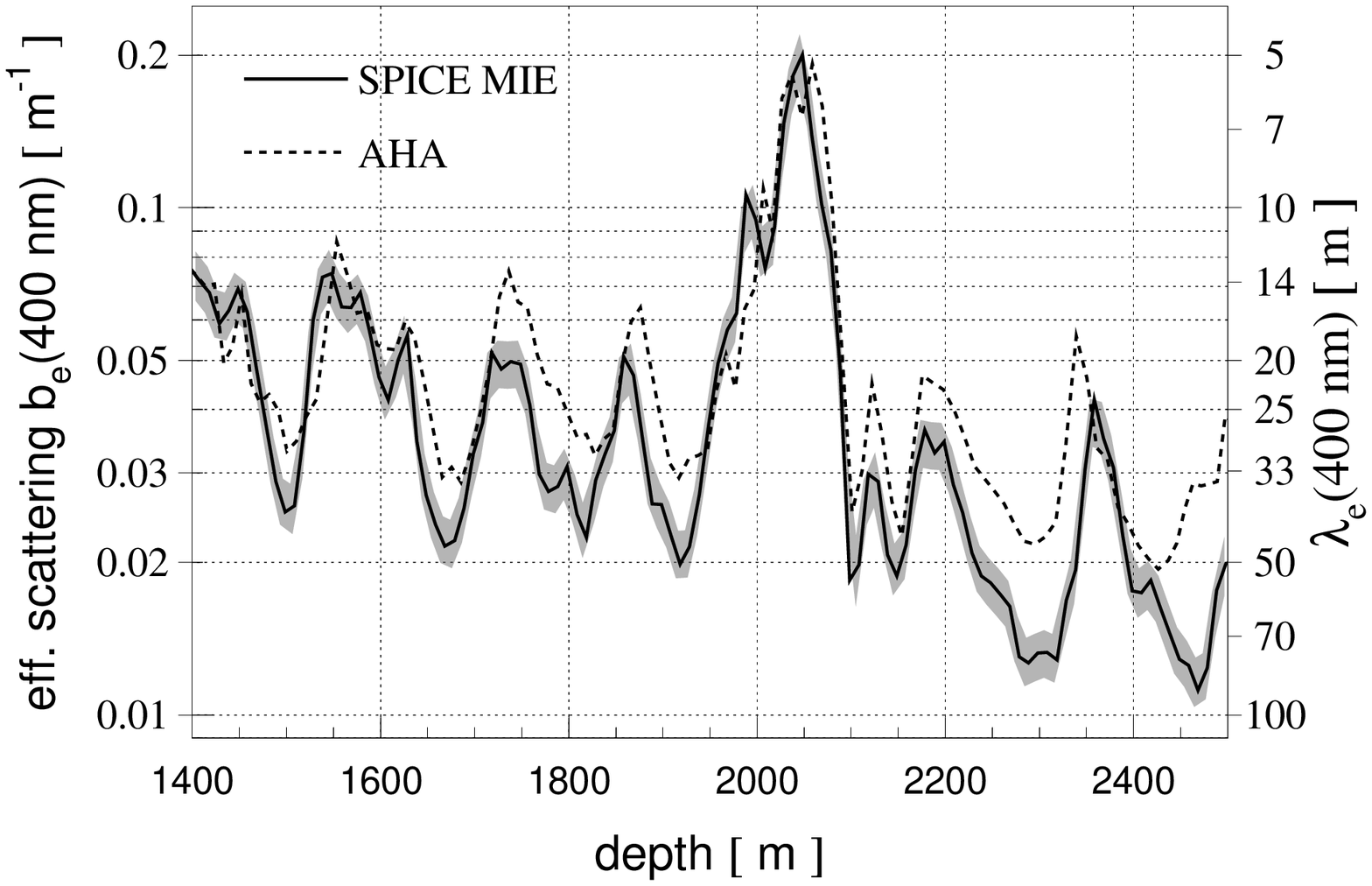,width=.75\textwidth}} \\
\mbox{\epsfig{file=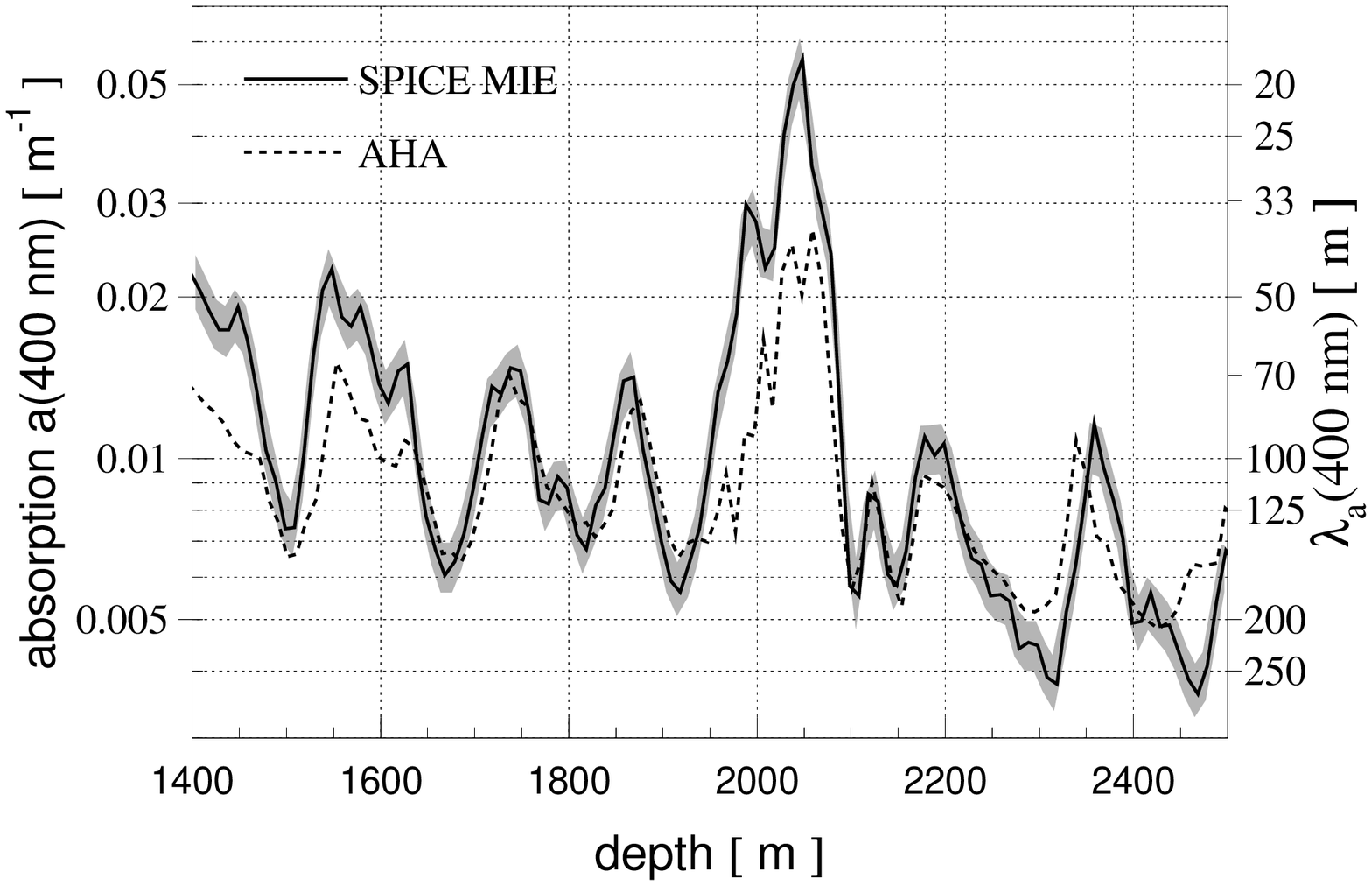,width=.75\textwidth}} \\
\end{tabular}
\parbox{.9\textwidth}{\caption{\label{xfit} Values of the effective scattering coefficient $b_e(400)$ and absorption coefficient $a(400)$ vs.\ depth for a converged solution are shown with a solid line. The range of values allowed by estimated uncertainties is indicated with a grey band around this line. The updated model of \cite{kurt} (AHA) is shown with a dashed line. The uncertainties of the AHA model at the AMANDA depths of $1730\pm 225$ m are roughly 5\% in $b_e$ and roughly 14\% in $a$. The scale and numbers to the right of each plot indicate the corresponding effective scattering $1/b_e$ and absorption $1/a$ lengths in [m]. }}
\end{center}\end{figure}

To study the precision of the reconstruction method, a set of flasher data was simulated with PPC (250 events for each of the 60 flashing DOMs). The agreement between the simulated and reconstructed ice properties is within 5\% at depths of the instrumented part of the detector (see Fig.\ \ref{sim}). Due to the dramatically lower number of recorded photons in the layer of ice containing the most dust, at around 2000~m (the {\it dust peak}), additional simulation was necessary to reconstruct the local ice properties: 250 events per flasher were used within the {\it dust peak}, whereas only 10 events per flasher were used elsewhere. Up to 250 simulated events per flasher were used to achieve the best possible precision of the final result shown in Fig.\ \ref{xfit}.

This verification approach was used to quantify the uncertainty in the measured values of $b_e(400)$ and $a(400)$ due to the lack of knowledge of the precise flasher output timing profile. Reconstructing the simulation, which used the 62 ns-wide rectangular shape of the flasher pulse, with a hypothesis that all photons are emitted simultaneously at flasher start time, leads to maximal systematic shifts in the obtained effective scattering and absorption coefficients of roughly 6.5\%.

Several pulse extraction methods, with and without correcting for PMT saturation (using the saturation model of \cite{pmt}), were tested for extracting photon hits from the flasher data, and the ice properties were reconstructed for each and compared. This provided an estimate of about 4\% for the uncertainty in the measured ice properties (effective scattering and absorption coefficients) due to detector calibration and pulse extraction (in waveforms of up to 1000 photoelectrons).
We note that reconstructing the data with the azimuthally symmetric vs.\ 6-fold ``star'' pattern of flasher LED light leads to no discernible difference in the resultant ice properties. Further, since the DOMs on the flashing string are not used in the fits, the difference between the ice properties reconstructed for nominal or hole ice angular sensitivity models is negligible.

Finally, the uncertainty due to statistical fluctuations in the sets simulated during the reconstruction procedure are estimated at roughly 5-7\%. This uncertainty could be reduced with more simulated events per flasher (at least 10 were simulated for each configuration, compared to 250 events present in data). However, given that the entire fitting procedure currently exceeds 10 days of calculation to produce a result, the number of simulated events represents a limiting constraint.

The effective scattering and absorption parameters of ice measured in this work are shown in Fig.\ \ref{xfit} with the $\pm 10$\% grey band corresponding to $\pm1\sigma$ combined statistical and systematic uncertainty at most depths (values of 6.5\%, 4\%, and $5-7$\% explained above are added in quadrature to result in a total uncertainty estimate of 9.7$\pm$0.6\%).
The uncertainties may be somewhat larger than this average value in the layers of dirtier ice, since many of the detected photons are likely to spend more of their travel time in the adjacent layers of cleaner ice (thus resulting in a weaker constraint of the properties of a dirtier ice layer).
The uncertainty increases beyond the shown band at depths outside the detector, above 1450~m and below 2450~m.

Figure \ref{xfit} also shows the AHA (Additionally Heterogeneous Absorption) model, based on the ice description of \cite{kurt}, extrapolated to cover the range of depths of IceCube and updated with a procedure enhancing the depth structure of the ice layers. The AHA model provided the ice description used for Monte Carlo simulations of IceCube data prior to this work.

\begin{figure}[!h]\begin{center}
\begin{tabular}{ccc}
\mbox{\epsfig{file=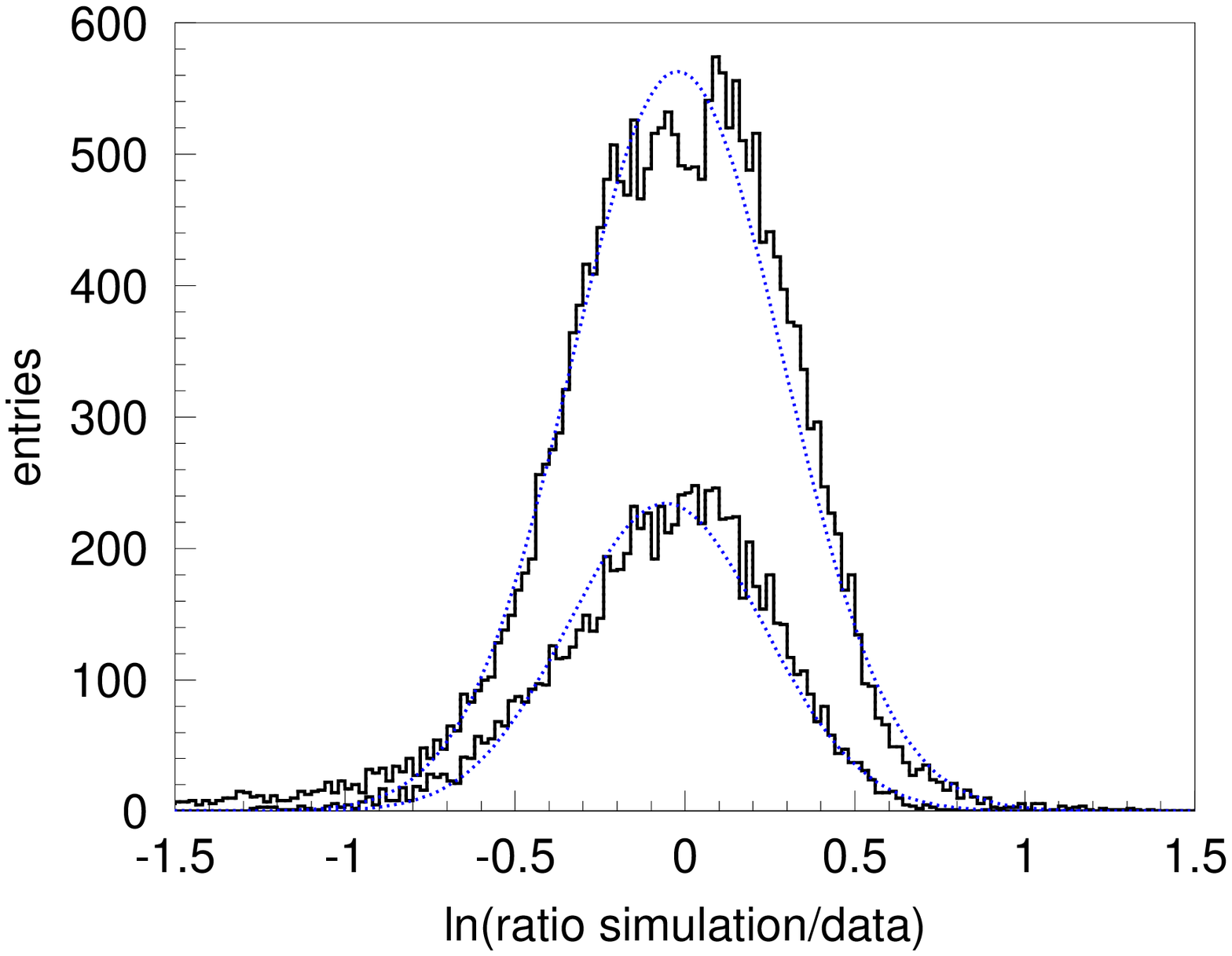,width=.5\textwidth}} \\
\end{tabular}
\parbox{.9\textwidth}{\caption{\label{ratio} Histogram of the ratio of the average charge per emitter-receiver pair in simulation to data. The lower histogram has one entry per pair, based on the averaged total charge. The upper one has one entry per time bin. Pairs and time bins are used only if the average charge in the data is more than 10 photoelectrons. 
%Ratio of simulation to data: lower histogram uses total (integrated) charges per event, while upper histogram uses charges binned in 25 ns time bins.
The widths of the fitted log-normal distributions are 0.29 and 0.31, respectively. }}
\end{center}\end{figure}

How well we fit the ice properties is limited by our ability to properly simulate all propagation and instrumental effects.
Not all effects are accounted for, as it appears, in the analysis presented here, despite the simplicity of the physics model involved.
In order to estimate the error in the description of the data with the fitted model,
we created a histogram (see Fig.\ \ref{ratio}) of the ratio of simulation to data for sufficiently large charges, minimizing statistical effects. 
The width of this histogram, estimated to be around 30\% of the received charge, represents the ``model error'' and enters the fit as a parameter in the likelihood function (see section \ref{llh}).

It is not possible to translate all of this model error into the uncertainties in the measured parameters since we can only estimate uncertainties from the known causes (e.g., by varying the parameters of the PMT saturation model). During our investigation of the discrepancy demonstrated in Fig.\ \ref{ratio}, we found evidence of different propagation properties of photons in different directions inside the detector.
%By introducing an anisotropy in the photon scattering function we could explain most of the observed variation in received charges in DOMs on strings around string 63. This effect has also been observed in flasher data from other locations in the detector as well as in the bulk of the normal IceCube data. It appears that a clear reduction in the fit error can be achieved, while the average effective scattering and absorption vs.\ depth remain within the error band of the values presented here. The effect is currently under investigation and will be reported in a coming paper. It is clear that none of the sources of systematic uncertainties considered above could lead to an uncertainty in measured ice parameters that is due to anisotropy.
Nevertheless, the uncertainties given above are still applicable to average (over all directions) values of effective scattering and absorption. The resulting situation compels us to report both the parameter uncertainties ($\sim$10\%), and the average model error ($\sim$30\%, when describing the flasher data as shown in Fig.\ \ref{ratio}).

\section{Comparison of full-detector simulation to muon data}
\label{comparison}
To investigate the accuracy of the resultant ice model in describing actual IceCube data, analyses were performed that compared experimental muon events with simulation. Atmospheric muons are a source of physics
events for IceCube but represent a background for neutrino analyses. In the 2008 40-string configuration, atmospheric muons triggered IceCube at a typical rate of 1 kHz, and therefore a large statistical data set was available for
comparisons between measured muon data and simulations of cosmic ray induced muons. The
simulations are based on the assumed propagation of optical Cherenkov photons through the ice but
also depend on assumptions that include the energy, multiplicity, and angular distribution of the muons.

The simulation chain begins with the production of atmospheric muons from cosmic
ray air showers using the CORSIKA software \cite{corsika}, followed by propagation of
the muons with muon Monte Carlo (MMC) \cite{mmc} and generation of photons according to a Cherenkov spectrum and their propagation with photonics \cite{effh} or PPC \cite{ppc}. Finally the photons are detected and digitized by the DOM simulator. To compare different
ice models and photon propagation techniques, only the parameters relevant for the photon
propagation are varied in simulation, while all other settings remain fixed.

\subsection{DeltaT distributions}
\begin{figure}[!h]
\centering
\includegraphics[width=.45\textwidth]{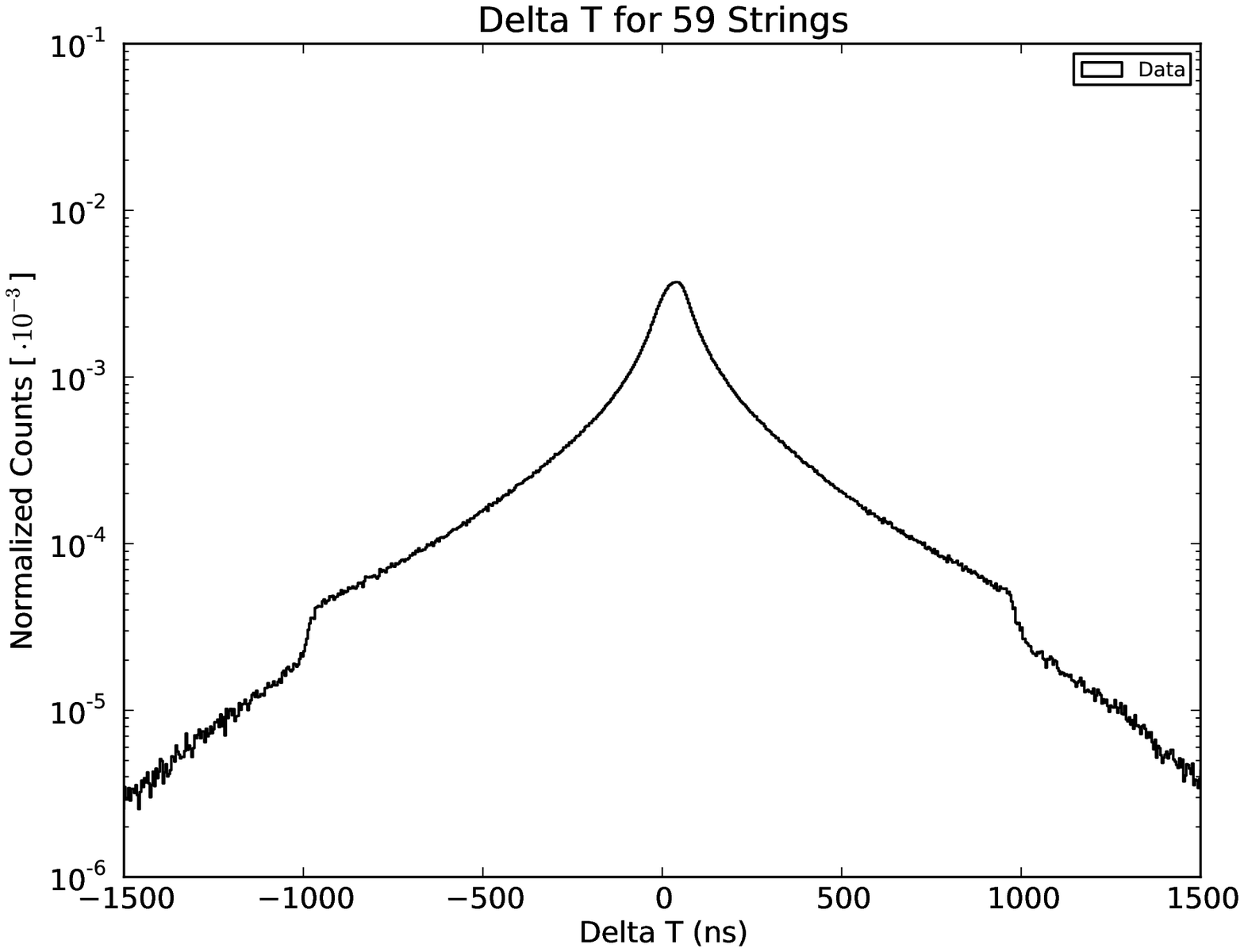}
\includegraphics[width=.45\textwidth]{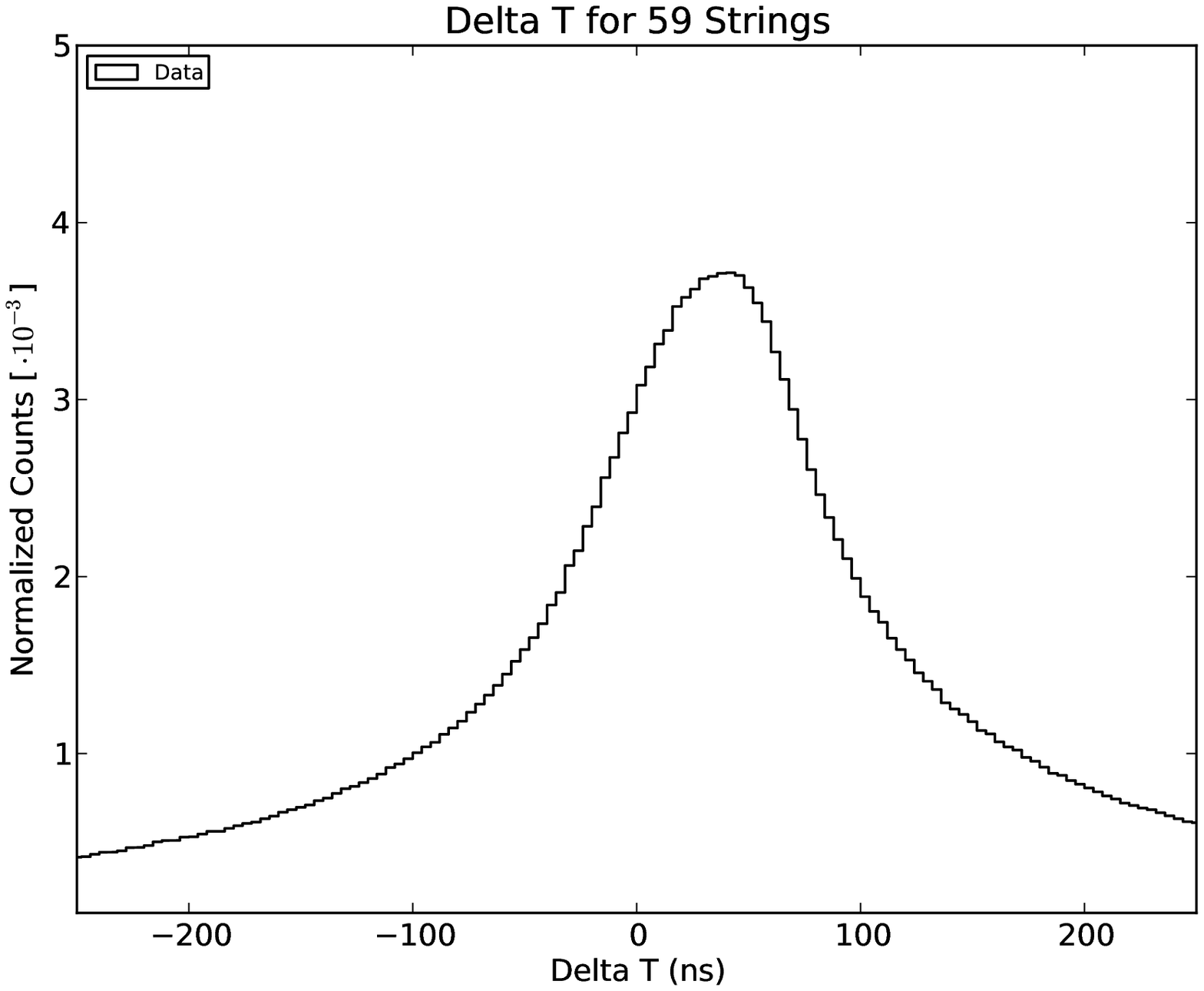}
\caption{\footnotesize{Left: DeltaT distribution for muon data.  The cutoff at $ \pm $ 1000 ns is due to the coincidence trigger window where data from a triggered DOM will only be read out if an adjacent or next-to-adjacent DOM also triggers within a time window of 1000 ns.  Right: A zoom of the peak of the distribution.  The peak is shifted towards positive times because it is dominated by direct photons from downgoing muons, which are detected first by the upper DOM and then the lower DOM.  The shift roughly corresponds to the muon flight time between DOMs.}}
\label{deltaT1}
\end{figure}

\begin{figure}[!h]
\centering
\includegraphics[width=.45\textwidth]{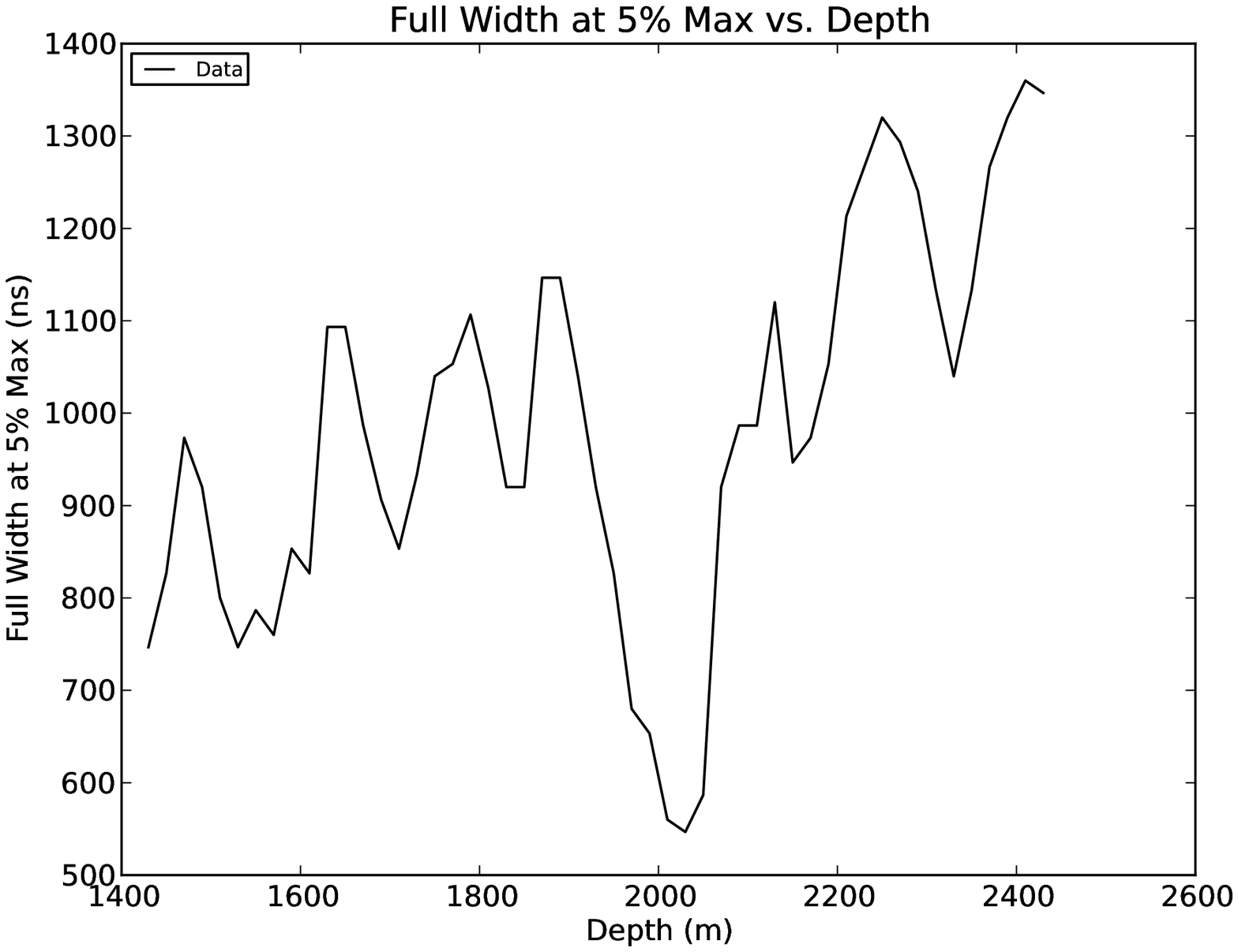}
\includegraphics[width=.45\textwidth]{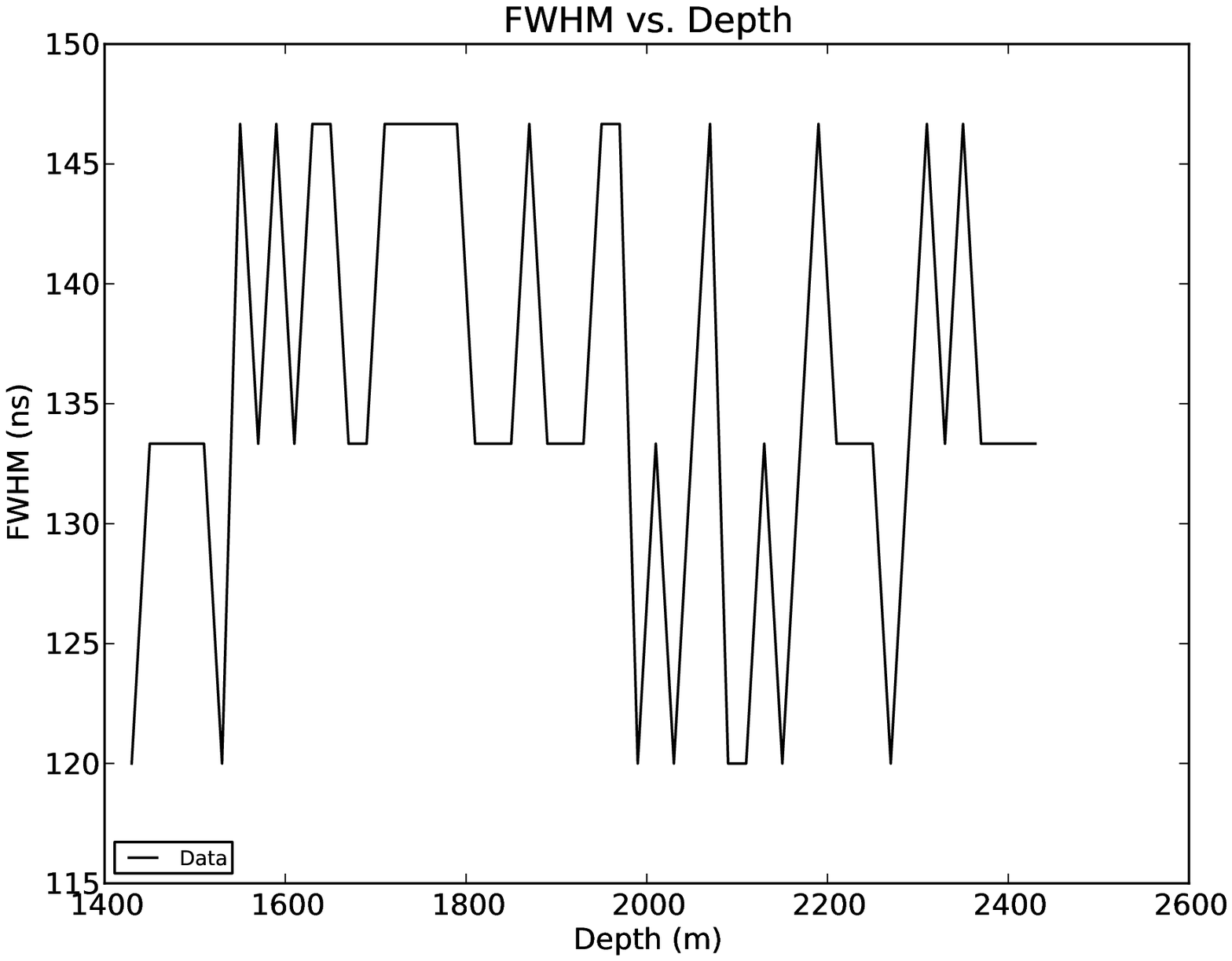}
\caption{\footnotesize{DeltaT distributions for DOMs binned in 20 m depths. Widths of the peaks and tails were extracted and plotted vs. depth for the entire detector.  Left: The full width at 5\% of the maximum, corresponding to the width of the tails, shows a strong depth dependence similar to the dust logger data and the derived scattering and absorption parameters.  Right: The full width at half maximum (FWHM) shows very little depth dependence. The FWHM was computed by multiplying the number of bins and the bin width, resulting in the discrete level structure in the plot.}}
\label{deltaT2}
\end{figure}

A relatively generic method to compare ice models and examine specific ice properties described here utilizes DeltaT distributions. DeltaT is defined as the time difference between first hits on adjacent DOMs on the same string. A positive DeltaT represents a photon that strikes the upper DOM followed by a photon strike of the DOM directly below. This method permits close investigation of basic photon timing information without requiring ice-model-dependent muon reconstruction techniques.  The distribution of DeltaT values for downgoing muon data taken with the 59-string detector configuration during September 2009 is shown in Fig.\ \ref{deltaT1}.  The tails of this distribution consist of relatively long-lived photons and contain information about the bulk ice properties, such as scattering and absorption.  On the other hand, the peak of the distribution consists of photons that travel from source to DOM with few scatters (i.e., ``direct'' photons), and is relatively invariant to the depth-dependent bulk ice properties.  Figure \ref{deltaT2} illustrates this relationship throughout all detector depths.

\begin{figure}[!h]
\centering
\includegraphics[width=.75\textwidth]{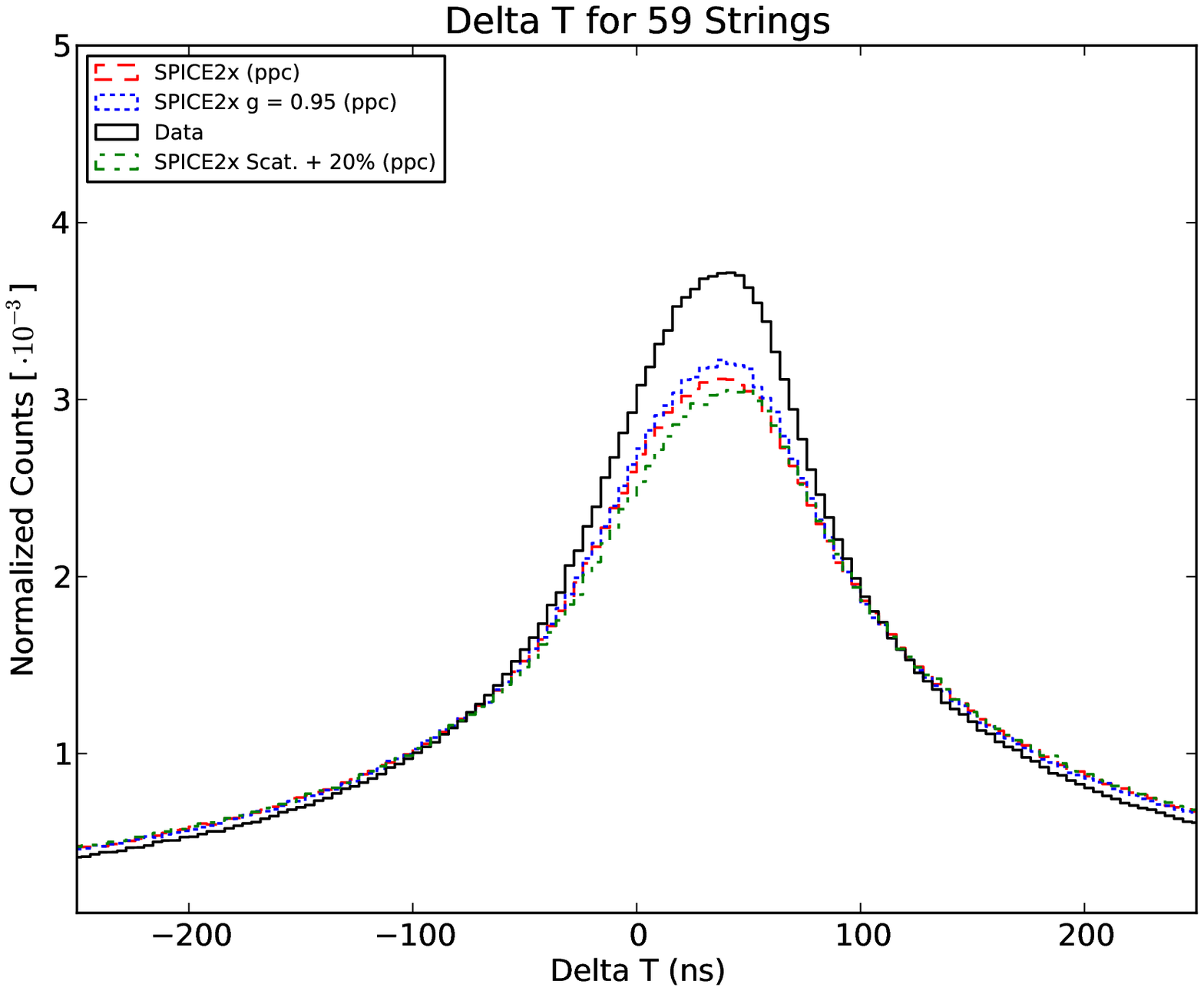}
\caption{\footnotesize{The peak region of the DeltaT distribution for the SPICE2x model shows a lack of direct photon hits compared to the data.  Neither increasing the amount of forward scattering (by setting g = 0.95) nor increasing the bulk ice scattering by 20\% significantly changes the peak height or shape.}}
\label{deltaT3a}
\end{figure}

\begin{figure}[!h]
\centering
\includegraphics[width=.75\textwidth]{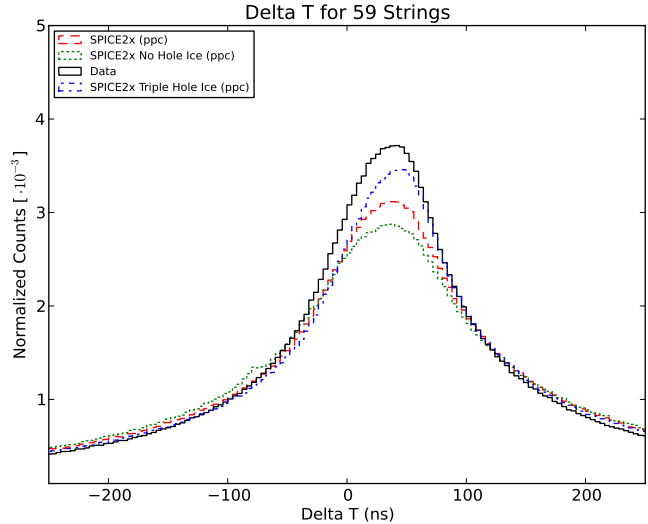}
\caption{\footnotesize{The peak region of the DeltaT distribution shows sensitivity to the hole ice description.  The hole ice is modeled as a vertical column of ice with a higher concentration of air bubbles, resulting in a local scattering length of 50 cm.  Simulations that assume no contribution to scattering due to hole ice and those with three times the nominal bubble concentration in the hole ice are shown.  The hole ice is thought to increase the number of direct photon hits because the increased scattering in the region of the hole ice causes more photons from downgoing muons to be locally up-scattered into the PMT (which faces downward in the DOM), effectively altering the angular sensitivity of the DOM.}}
\label{deltaT3b}
\end{figure}

\begin{figure}[!h]
\centering
\includegraphics[width=.75\textwidth]{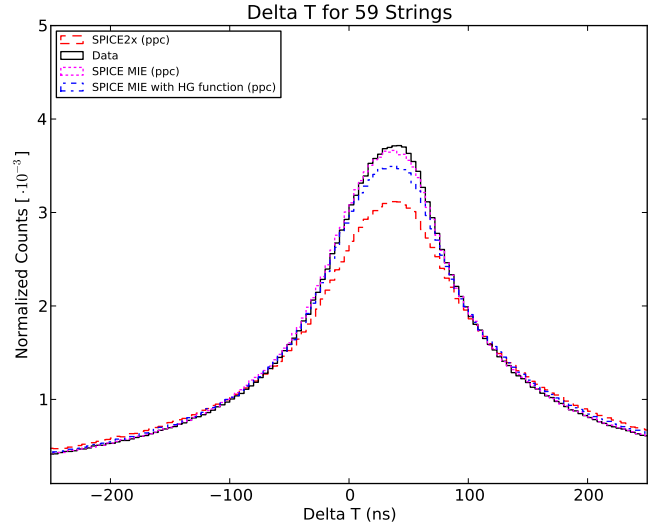}
\caption{\footnotesize{The peak region of the DeltaT distribution for SPICE Mie, comparing the full model ($f_{SL}=0.45$) to the model with only the HG scattering function (i.e., setting $ f_{SL}=0 $).  The observed effect is thought to be caused by the higher probability of photons scattering through intermediate angles of $20^{\circ}$-$40^{\circ}$.  Even though the typical muon-to-DOM distance is small compared to the effective scattering length, photons are more likely to scatter at larger angles and therefore to be detected.}}
\label{deltaT3c}
\end{figure}

\begin{figure}[!h]
\centering
\includegraphics[width=.75\textwidth]{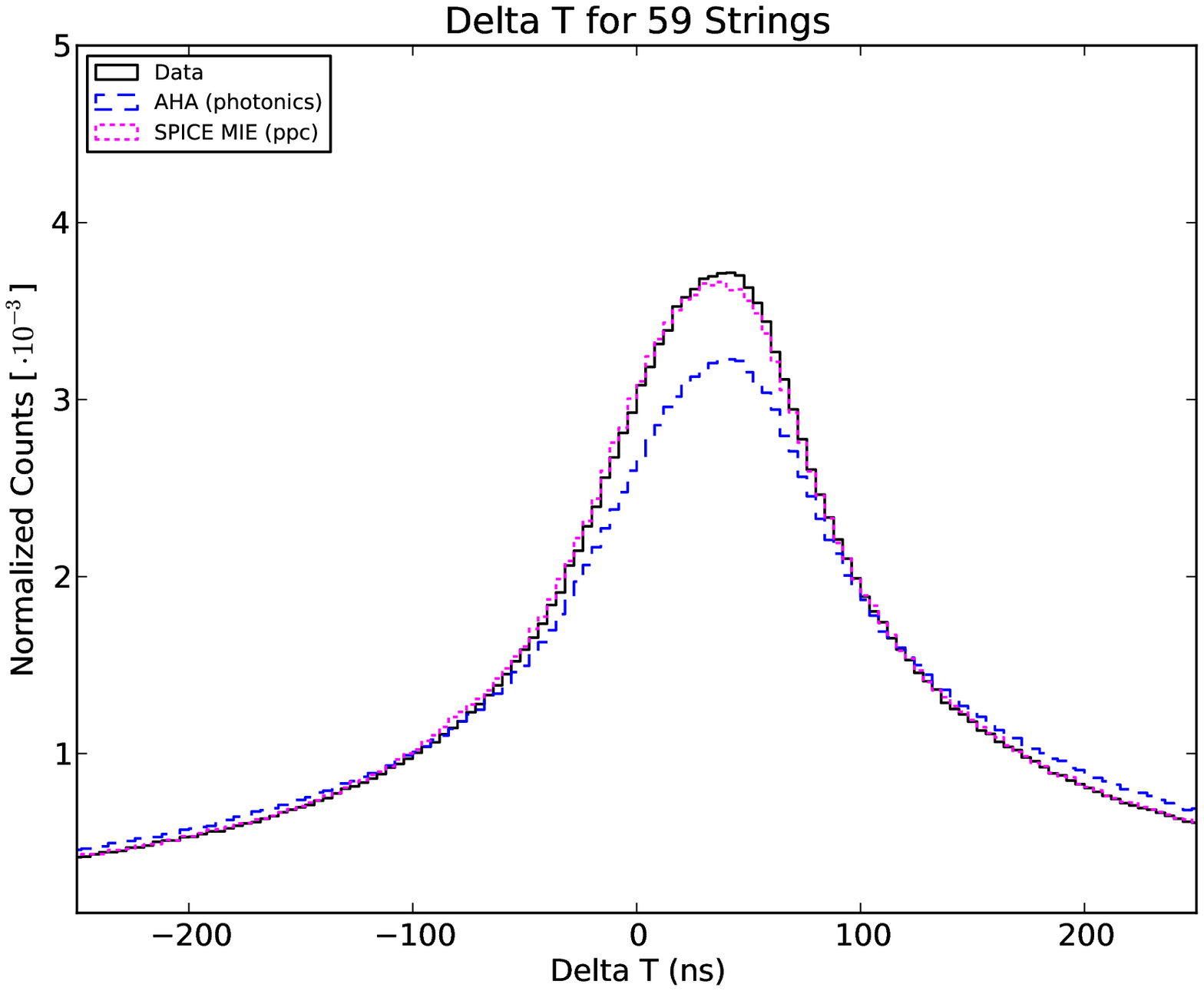}
\caption{\footnotesize{The peak region of the DeltaT distribution comparing the final SPICE Mie fit result to the previous AHA model and the muon data. The fit to the data is significantly improved with the SPICE Mie model.}}
\label{deltaT3d}
\end{figure}

Full-detector Monte Carlo simulation was generated for different ice parameters to examine their effects on the shape and height of the peak in the DeltaT distribution.   Figures \ref{deltaT3a}-\ref{deltaT3d} show the peak shape for data and various simulation models. The description of the ice denoted as SPICE2x was an intermediate model in this analysis, and is characterized by similar scattering and absorption lengths to those of the SPICE Mie model, which is the final result. In SPICE2x, a Henyey-Greenstein (HG) scattering function is used instead of a linear combination of the HG and SL functions.  Additionally, SPICE2x has an average scattering angle of $g = \langle \cos\theta\rangle = 0.8$ instead of 0.9 (used in the final result), and lacks the global flasher time offset parameter used in the fit of SPICE Mie.  In all of the permutations of the ice properties examined, the only parameters that significantly changed the shape of the peak were the hole ice scattering, scattering function composition, and the time offset parameter.
%  Figure \ref{deltaT3d} shows the DeltaT distribution comparing the resultant final fit of SPICE Mie to the previous AHA model.

\subsection{Event geometry and time residuals}

The simulation data for this study were produced for the IceCube detector in its 40-string
configuration and is compared to data taken in
August 2008. This corresponds to roughly 10\% of the yearly experimental data of IceCube.

\begin{figure}[!h]\begin{center}
\begin{tabular}{ccc}
\mbox{\epsfig{file=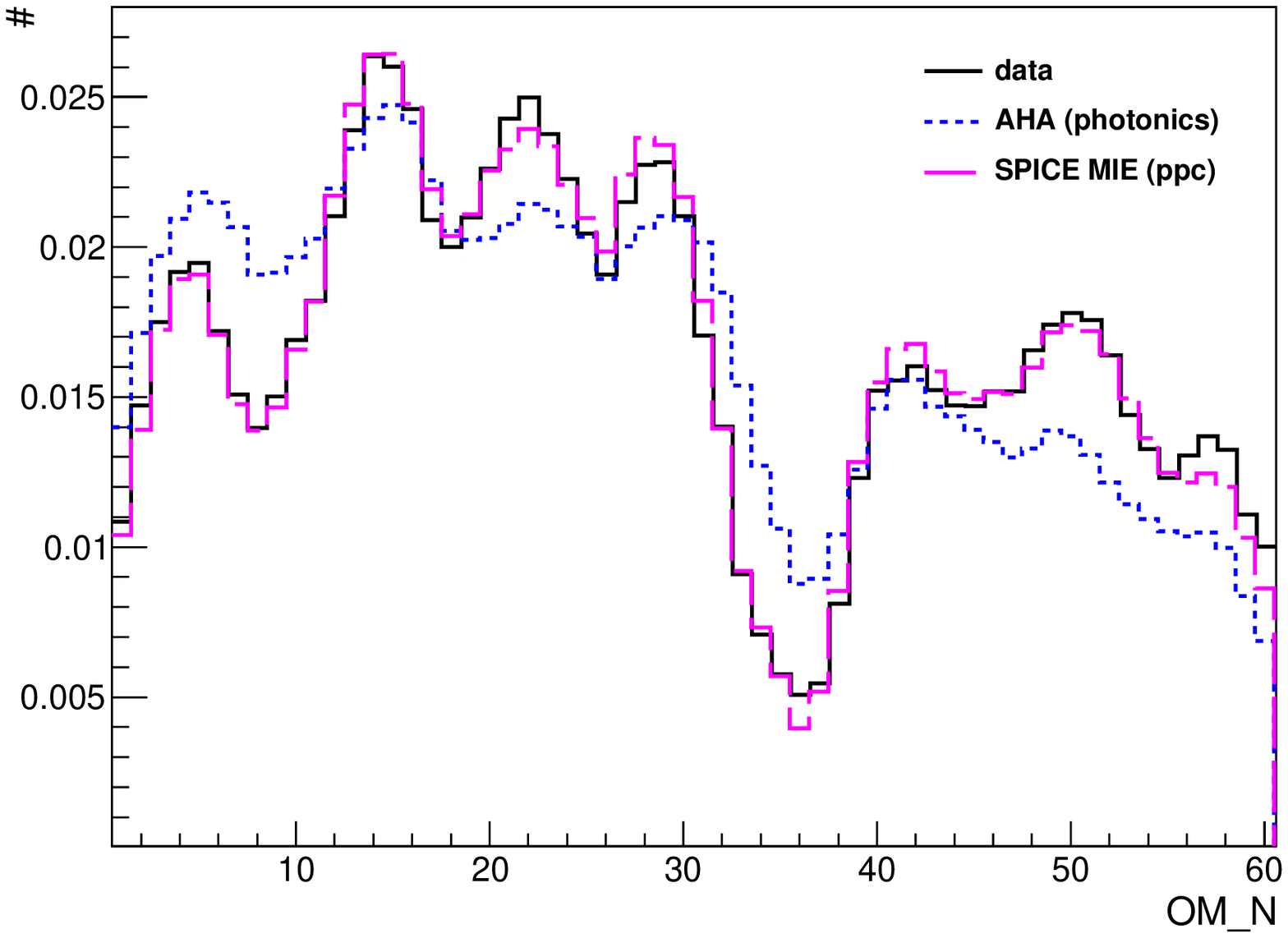,width=.46\textwidth}} & \ & \mbox{\epsfig{file=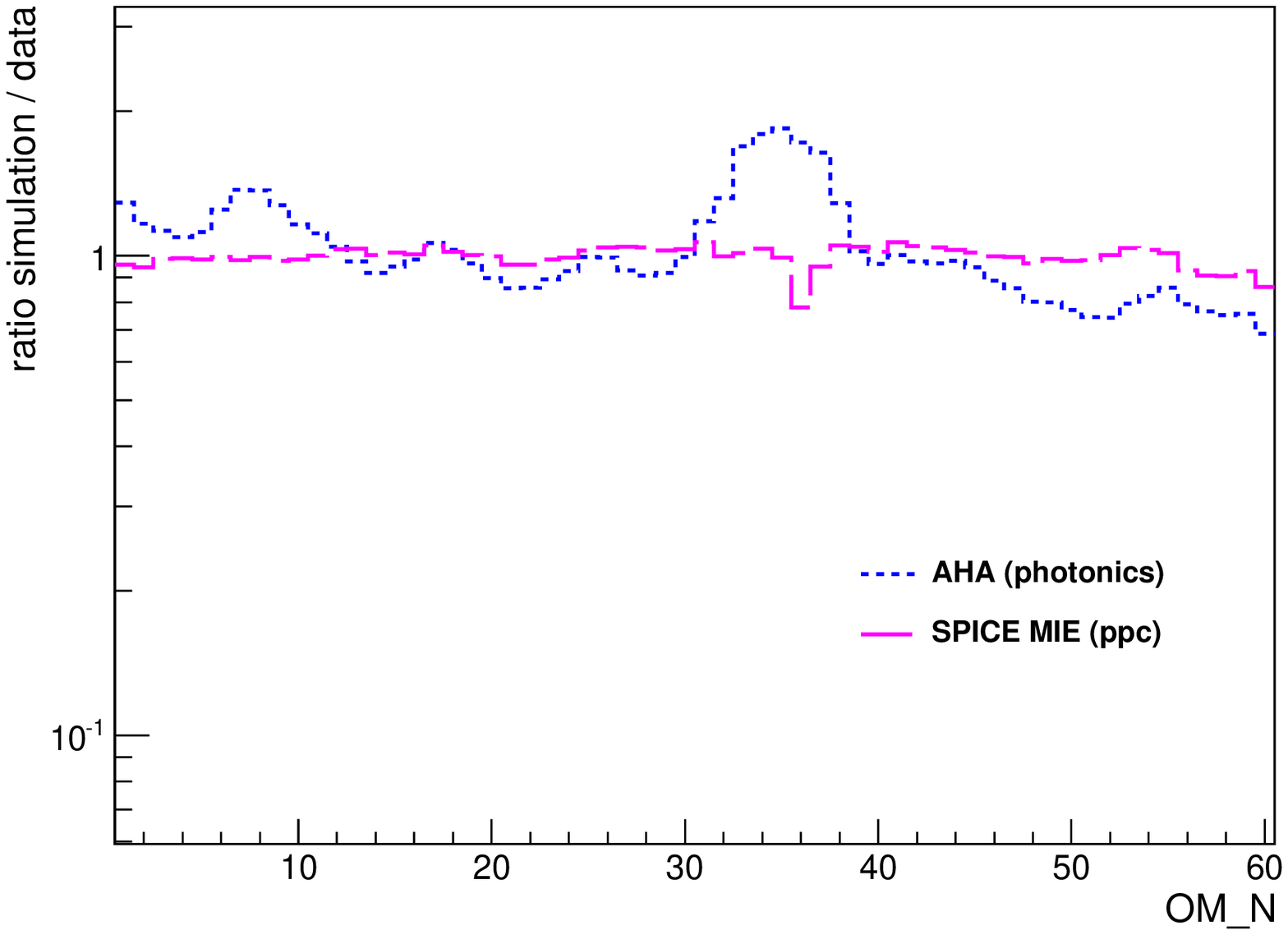,width=.45\textwidth}} \\
\end{tabular}
\parbox{.9\textwidth}{\caption{\label{muo1} Distribution of hit DOMs. The label ``OM\_N'' is the number of the DOM on the string, ranging from 1 at the top of the detector to 60 at the bottom. The curves are normalized to one event for a better comparison of the shape. The plot on the right shows the ratio between simulation and data.}}
\end{center}\end{figure}

\begin{figure}[!h]\begin{center}
\begin{tabular}{ccc}
\mbox{\epsfig{file=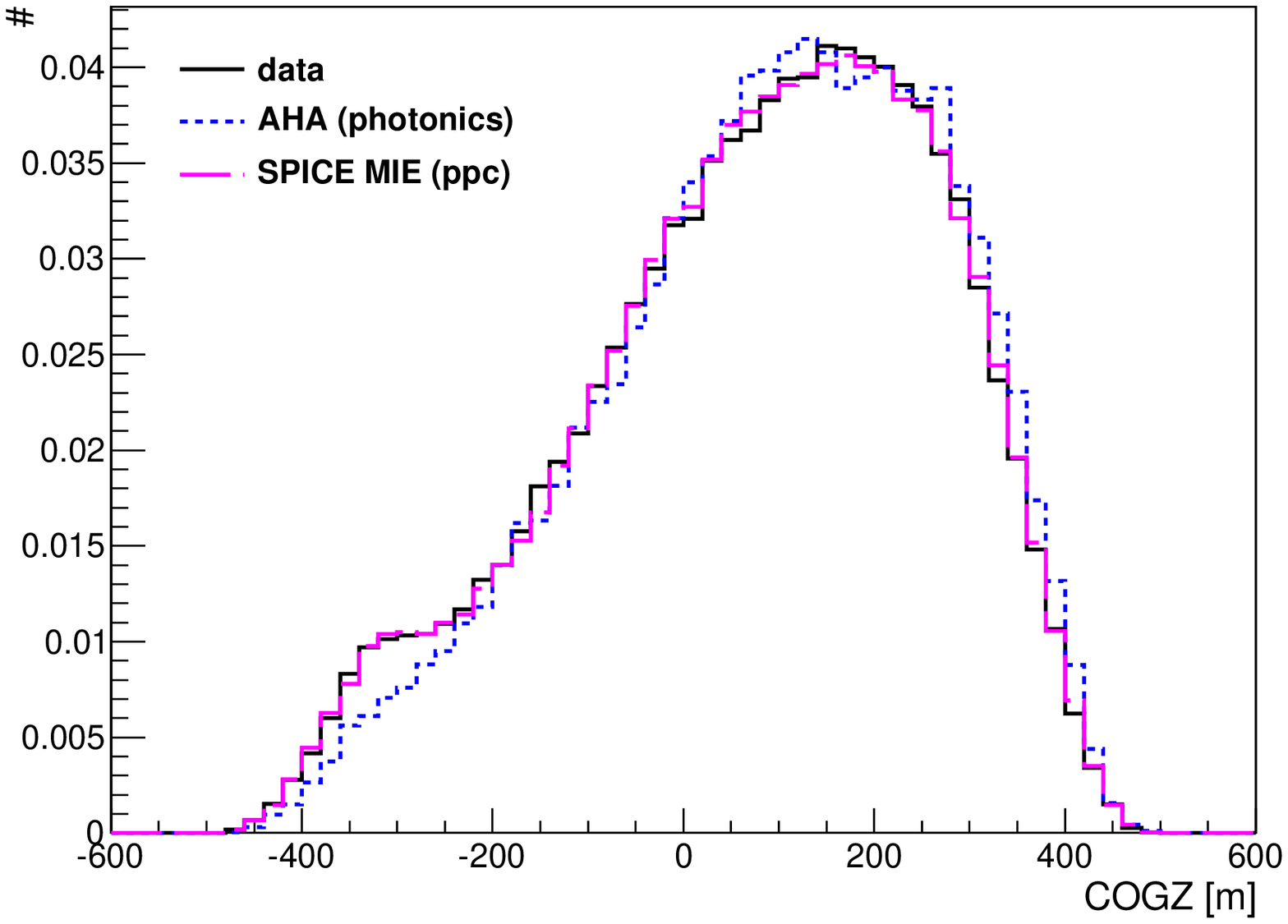,width=.46\textwidth}} & \ & \mbox{\epsfig{file=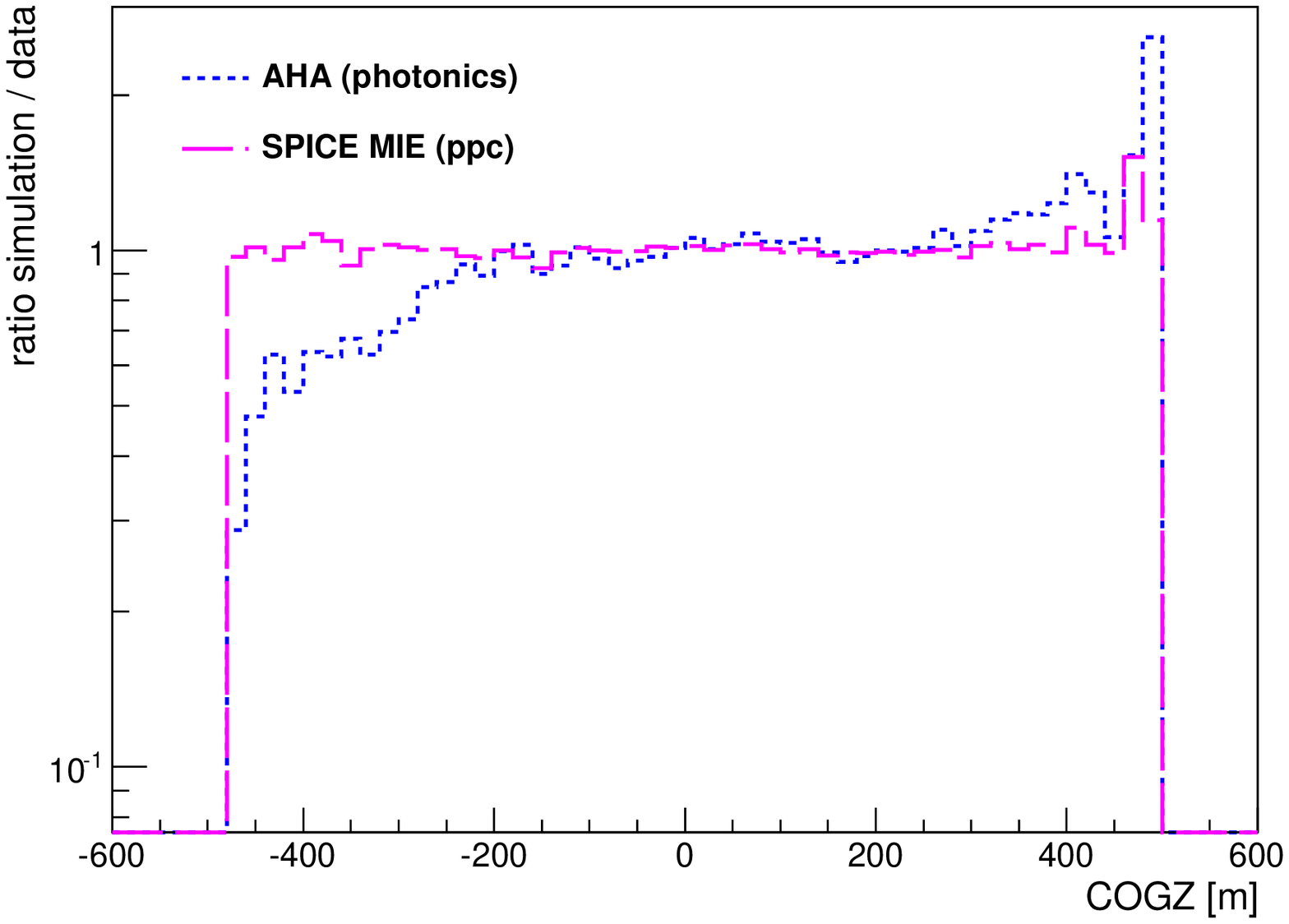,width=.45\textwidth}} \\
\end{tabular}
\parbox{.9\textwidth}{\caption{\label{muo2} Distribution of the $z$-coordinate of the center of gravity of all hit coordinates for each event. The curves are normalized to one event for a better comparison of the shape. The plot on the right shows the ratio between simulation and data. }}
\end{center}\end{figure}

\begin{figure}[!h]\begin{center}
\begin{tabular}{ccc}
\mbox{\epsfig{file=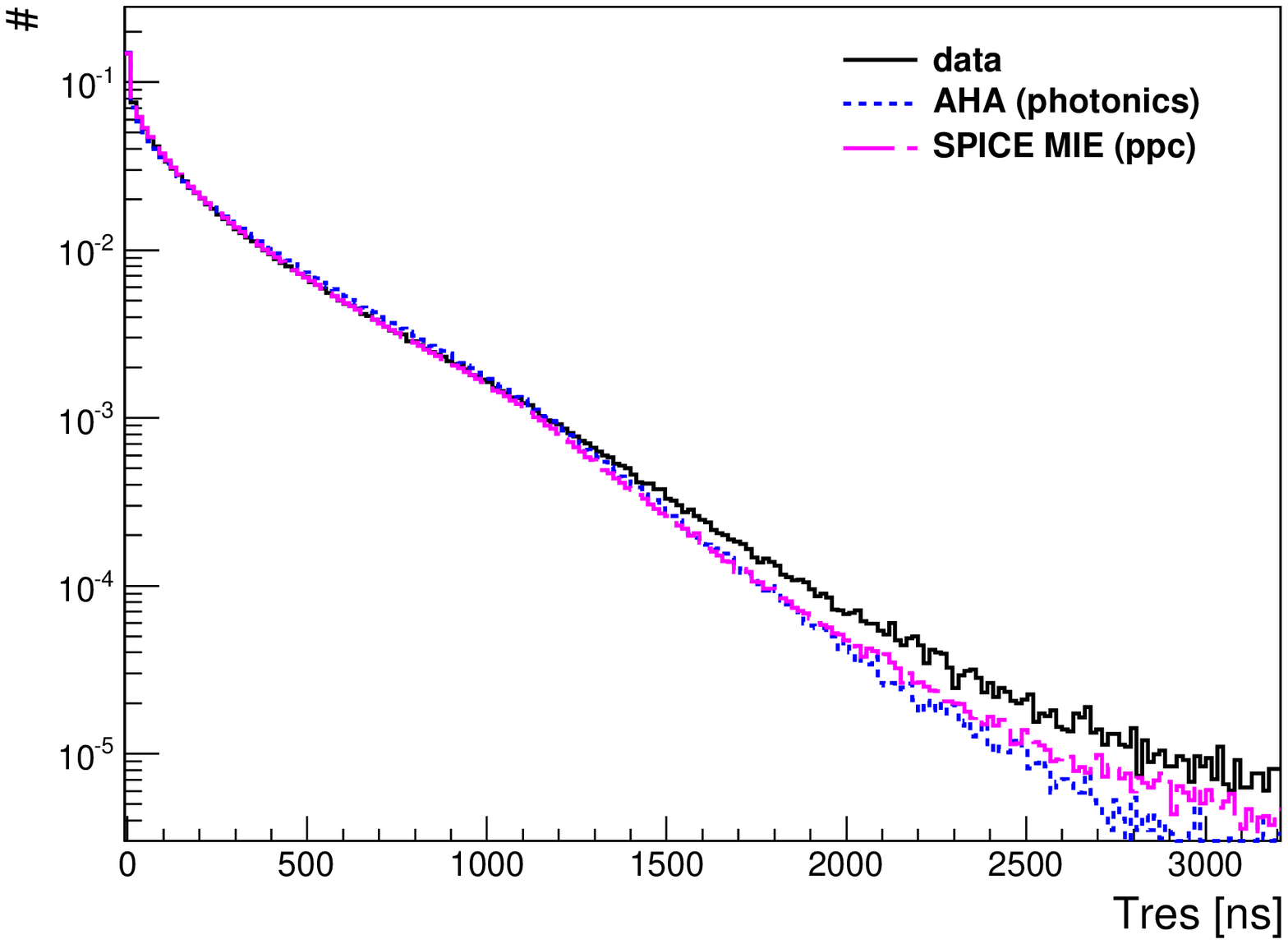,width=.46\textwidth}} & \ & \mbox{\epsfig{file=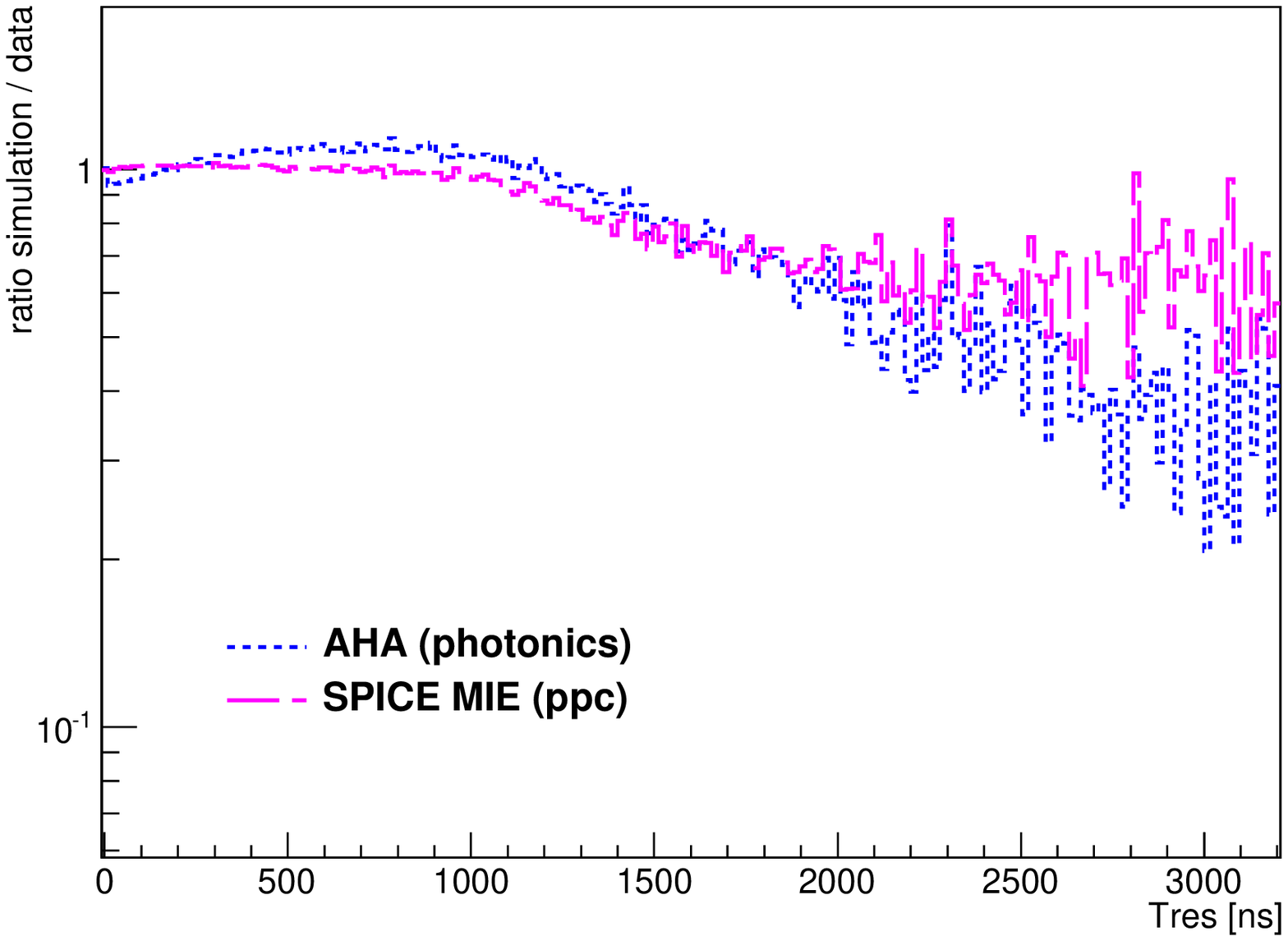,width=.45\textwidth}} \\
\end{tabular}
\parbox{.9\textwidth}{\caption{\label{tres} Distribution of the time residuals: time delay due to scattering of photons arriving from the reconstructed muon track in data and simulation. The curves are normalized to one event for a better comparison of the shape. The plot on the right shows the ratio between simulation and data. }}
\end{center}\end{figure}

For a generic comparison, it is preferable to use an unbiased data sample.
For such purposes,
IceCube operates a Minimum Bias filter stream that selects every event that was recorded by
the DAQ independently of the satisfied trigger condition. A prescale factor of 2000 was applied to data events
to comply with bandwidth requirements before sending data north via satellite. This analysis used events that passed a DOM multiplicity condition  of at least 8 DOMs within a 5 $\mu$s time window that register a hit in coincidence with one of their vertical neighbors (within 1 $\mu$s).
From this data
stream, events that had sufficient recorded information to perform reconstructions of reasonable accuracy  were selected.
The selection criteria were based on the zenith angle ($\theta <90^{\circ}$) and the likelihood minimum of
the standard angular fit \cite{amareco} divided by the degrees of freedom (reduced log-likelihood,  $rlogl < 8$). The resulting median angular resolution of this event sample was
better than 2$^{\circ}$ with a passing rate of roughly 15\% of the initially recorded data.
The comparisons shown in Figs.\ \ref{muo1}-\ref{tres} are based on 130 million events. The absolute normalization
between experimental data and simulation was affected by large uncertainties, but for the purpose of this
analysis all distributions were normalized to unity, and the focus was placed on differences in shape between the curves.

A basic test to examine the influence of ice parameters on the simulation is to compare depth-dependent variables since the optical ice properties characteristically vary with depth. Figure
\ref{muo1} shows the distribution of hit DOMs. The peaks correspond to regions with clearer ice, and the valleys indicate ice containing more dust. The simulation that uses a combination of the SPICE Mie ice model and the PPC photon propagator shows a significant improvement in agreement with the experimental data. The ratio between Monte Carlo and experimental data histograms is almost flat,
except for a dip around DOM 36, which is a region of high dust concentrations and therefore poor statistics. Figure \ref{muo2} shows the distribution of the $z$-coordinate of the center of gravity of all hit DOMs for
each event. A marked improvement in the agreement between
experiment and simulations, in particular in the deep ice, is observed for the SPICE Mie and PPC Monte Carlo.

Similar to the timing distributions for the flasher events that were used to extract the ice properties,
the distribution of arrival times for Cherenkov light from muons can be investigated. Here we
study the distribution of time residuals, which are calculated by subtracting the expected geometrical arrival time for unscattered Cherenkov photons (based on the estimate of the reconstructed track) from the actual arrival time of the detected photons. If the track is reconstructed accurately, the time residual is caused by scattering. The slope
of the time residual distribution is strongly correlated with the optical absorption length.
The new simulation shows an improved overall description of the experimentally observed timing residuals, see Fig.\ \ref{tres}. The plot of the ratio between Monte Carlo and experimental data histograms shows an almost flat behavior for the most relevant time interval up to 1 $\mu$s. Note that the distribution is summed over depth so discrepancies at specific depths may remain.

\section{Conclusions and outlook}
\label{results}
The precise modeling of the optical properties of the South Pole ice is crucial to the analysis of IceCube data. The scattering and absorption coefficients of ice (averaged in 10~m depth bins) were obtained from a fit to the in-situ light source data collected in 2008. The result is shown in Fig.\ \ref{xfit} and also presented in \apname\ref{tresults}. The sum of the statistical and systematic uncertainties in the measured values of the effective scattering and absorption coefficients inside the instrumented volume of the IceCube detector was estimated to be less than 10\%.

This analysis builds on the concepts developed in \cite{kurt}, and relies on the wavelength dependence determined there. Unlike in \cite{kurt}, this analysis uses a global fit to simultaneously describe all of the light source data.  It also uses significantly more data than \cite{kurt}, both in terms of the number of registered photons and the number of emitter-receiver pairs.
%While some of the concepts developed in \cite{kurt} were used implicitly in this analysis, the main difference is the use of a global fit that was designed to simultaneously describe all of the applied light source data. Statistically, this work uses significantly more data compared to \cite{kurt}, both in the number of registered photons and, in particular, the number of combinations of emitters and receivers.

The high quality of the fit was ensured by careful selection of the likelihood function that quantified the differences between data and simulation within a single model of statistical and systematic uncertainties. In the course of the investigation we found that determining the shape of the scattering function and incorporating the ice layer tilt was necessary to achieve better model agreement with the data.

We are aware of some systematics issues that are not yet entirely understood (and will be the subject of further studies). One notable omission from this work is the direct simulation of the photon propagation in the columns of ice refrozen around the IceCube strings. A study of the slight anisotropy hinted at in this report (section \ref{errors}) is the subject of a forthcoming publication. Additionally, we have not yet analyzed the data from the LEDs with wavelengths other than 400~nm (which were installed during the final IceCube deployment season in 2010). Thus, the new ice model presented here relies on the previously-established wavelength dependence of effective scattering and absorption coefficients. However, we are encouraged by the significantly improved agreement between data and simulation when using the new ice mode obtained in this analysis.

\appendix

\section{Photon Propagation Code}
\label{ppc}

Four different versions of the program (available from \cite{ppc}) were written: one in C++, another in assembly (for the 32-bit i686 with SSE2 architecture), and a version that employs the NVIDIA GPUs (graphics processing units) via the CUDA programming interface \cite{cuda}. The fourth version uses OpenCL \cite{cuda}, supporting both NVIDIA and AMD GPUs, and also multi-CPU environments. The relative performance of these different implementations (for simulating both flashers and Cherenkov light from muons) is compared in Table \ref{cmp}.

\begin{table}[!h]\begin{center}
\begin{tabular}{|ccccc|}
\hline
& C++ & Assembly & CUDA & OpenCL \\
flasher & 1.00 & 1.25 & 147 & 105 \\
muon & 1.00 & 1.37 & 157 & 122 \\
\hline
\end{tabular}
\caption{\label{cmp}Speedup factor of different implementations of PPC compared to the C++ version. C++ and Assembly code was run on 1 core of Intel i7 920 (2.66 GHz) CPU. CUDA and OpenCL code was run on 1 GPU of NVidia GTX 295 video card. }
\end{center}\end{table}

The writing of the GPU version of PPC was prompted by a similar project called i3mcml \cite{mcml}, which showed that acceleration by factor of 100 or more compared to the CPU-only version was possible. After demonstrating impeccable agreement between test simulation sets made with the C++, assembly, and GPU implementations of PPC, and with i3mcml, the CUDA version of PPC was chosen for the analysis of this work on a GPU-enabled computer with i7 920 (2.67 GHz) CPU and 3 GTX 295 NVIDIA cards (6 GPUs).

\section{Muon and cascade light production}
\label{light}

The light yield of the muon and all of its secondaries (ionization losses and delta electrons, bremsstrahlung, electron pair production, and photonuclear interaction \cite{mmc}) with energies below 500 MeV is parameterized in \cite{wiebusch}\footnote{The formula 7.97 contains a typo; however, the caption within Fig.\ 7.56 (B) is correct, with LOG(E) understood as $\ln(E)\equiv\log_e(E)$.} by substituting the length $dl$ of the Cherenkov-light-emitting segment of a {\it bare} muon of energy $E$ with the ``effective length''
\[dl_{\rm eff}=dl \cdot \left( 1.172+0.0324\cdot\log_e(E \mbox{ [GeV]}) \right).\]
The parameterization given above was used in the muon studies of section \ref{comparison}. However, we are aware of an updated parameterization of \cite{newwiebusch} and list it here for completeness:
\[dl_{\rm eff}=dl \cdot \left( 1.188+0.0206\cdot\log_e(E \mbox{ [GeV]}) \right).\]
The light yield of cascades is also parameterized in \cite{wiebusch} via use of the ``effective length'':
\bal
dl_{\rm eff}=0.894 \cdot 4.889/\rho \mbox{ } \mbox{ m/GeV} \cdot E \mbox{ [GeV]} & \quad \mbox{for electromagnetic cascades}\\
dl_{\rm eff}=0.860 \cdot 4.076/\rho \mbox{ } \mbox{ m/GeV} \cdot E \mbox{ [GeV]} & \quad \mbox{for hadronic cascades}.
\eal
These formulae were derived for muons in water, but are given here for propagation in ice ($\rho=0.9216$ is the ratio of the densities of ice\footnote{Taken at the center of IceCube (depth of 1950 m, temperature $-30.4^\circ$ C); cf.\ $\rho=0.9167$ at 1 atm.\ at $0^\circ$ C.} and water). This work relies on newer parameterization of the cascade light yield of \cite{marek}\footnote{The axis labels in Fig.\ 3.2 are correct; formula 3.4 needs to be corrected as in this text.}:
\bal
dl_{\rm eff}=5.21 \mbox{ m/GeV} \cdot 0.924/\rho \cdot E \mbox{ [GeV]} & \quad \mbox{for electromagnetic cascades} \\
dl_{\rm eff}=F \cdot 5.21 \mbox{ m/GeV} \cdot 0.924/\rho \cdot E \mbox{ [GeV]} & \quad \mbox{for hadronic cascades}.
\eal
Here F is a ratio of the effective track length of the hadronic to electromagnetic cascades of the same energy $E$. It is approximated with a gaussian distribution with the mean $\langle F \rangle$ and width $\sigma_F$:
\bal
\langle F \rangle = & 1-(E\mbox{ [GeV]}/E_0)^{-m}\cdot(1-f_0), \quad & E_0=0.399, \quad m=0.130, \quad f_0=0.467,\\
\sigma_F = & \langle F \rangle\cdot \delta_0 \cdot \log_{10}(E\mbox{ [GeV]})^{-\gamma}, \quad & \delta_0=0.379, \quad \gamma=1.160.
\eal

In order to properly account for the longitudinal development of the cascade, the distance from the start of the cascade to the point of photon emission is sampled from the following distribution (ignoring the LPM elongation) \cite{wiebusch}:
\[l=L_{rad} \cdot \Gamma(a)/b, \quad L_{rad}=35.8\mbox{ cm}/\rho,\]
where $\Gamma(a)$ is a gamma distribution with the shape parameter $a$. Parameters $a$ and $b$ are given by:
\bal
a=2.03+0.604\cdot\log_e(E), \quad b=0.633 & \quad \mbox{for electromagnetic cascades} \\
a=1.49+0.359\cdot\log_e(E), \quad b=0.772 & \quad \mbox{for hadronic cascades}.
\eal

\begin{figure}[!h]\begin{center}
\begin{tabular}{c}
\mbox{\epsfig{file=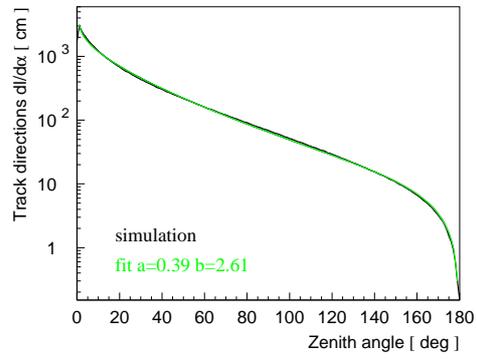,width=.45\textwidth}} \\
\end{tabular}
\parbox{.9\textwidth}{\caption{\label{fit} Fit to the angular track-length distribution for 100 GeV electron cascades. The simulation line in black is taken from figure 7.44 of \cite{wiebusch}, the fit in green is to the function given in this text. }}
\end{center}\end{figure}

All photons are emitted strictly at the Cherenkov angle with respect to the emitting track segment. These, except for the {\it bare} muon itself, are assumed to be distributed according to
\[dl/dx \sim \exp(-b\cdot x^a) \cdot x^{a-1}, \quad \mbox{with} \quad x=1-\cos(\theta).\]
The coefficients $a=0.39$ and $b=2.61$ were fitted to a distribution of 100 GeV electron cascades from \cite{wiebusch} (see Fig.\ \ref{fit}) and are fairly constant with energy and are used to describe the hadronic cascades as well.

\section{Table of results}
\label{tresults}

\begin{table}[!h]\begin{center}
\begin{tabular}[!h]{|c|c|c|}
\begin{tabular}{ccc}
depth, [m] & $1/b_e$, [m] & $1/a$, [m] \\
\hline
1398.4 & 13.2 & 45.1 \\ 
1408.4 & 14.0 & 48.6 \\ 
1418.4 & 14.7 & 53.2 \\ 
1428.4 & 17.0 & 57.6 \\ 
1438.4 & 16.0 & 57.6 \\ 
1448.4 & 14.4 & 52.2 \\ 
1458.4 & 16.0 & 60.1 \\ 
1468.4 & 20.8 & 74.6 \\ 
1478.4 & 26.7 & 96.6 \\ 
1488.4 & 34.7 & 110.5 \\ 
1498.4 & 39.7 & 135.6 \\ 
1508.5 & 38.7 & 134.7 \\ 
1518.6 & 27.8 & 98.2 \\ 
1528.7 & 16.6 & 64.7 \\ 
1538.8 & 13.7 & 48.5 \\ 
1548.7 & 13.5 & 44.3 \\ 
1558.7 & 15.7 & 54.4 \\ 
1568.5 & 15.7 & 56.7 \\ 
1578.5 & 14.7 & 52.1 \\ 
1588.5 & 17.6 & 60.7 \\ 
1598.5 & 21.6 & 72.7 \\ 
1608.5 & 24.0 & 78.9 \\ 
1618.5 & 20.0 & 68.7 \\ 
1628.5 & 17.8 & 66.6 \\ 
1638.5 & 28.9 & 100.0 \\ 
1648.4 & 36.9 & 128.6 \\ 
1658.4 & 42.1 & 148.2 \\ 
1668.4 & 46.5 & 165.7 \\ 
1678.5 & 45.4 & 156.0 \\ 
1688.5 & 39.1 & 138.5 \\ 
1698.5 & 30.6 & 113.9 \\ 
1708.5 & 26.5 & 90.2 \\ 
1718.5 & 19.3 & 73.5 \\ 
1728.5 & 20.8 & 75.9 \\ 
1738.5 & 20.1 & 67.8 \\ 
1748.5 & 20.3 & 68.6 \\ 

\end{tabular} &
\begin{tabular}{ccc}
1758.5 & 24.5 & 83.8 \\ 
1768.5 & 33.5 & 119.5 \\ 
1778.5 & 36.2 & 121.6 \\ 
1788.5 & 35.4 & 108.3 \\ 
1798.5 & 32.3 & 113.4 \\ 
1808.5 & 40.2 & 139.1 \\ 
1818.4 & 44.7 & 148.1 \\ 
1828.4 & 34.5 & 122.8 \\ 
1838.4 & 30.6 & 113.8 \\ 
1848.4 & 27.5 & 89.9 \\ 
1858.4 & 19.7 & 71.7 \\ 
1868.5 & 21.4 & 70.6 \\ 
1878.5 & 28.8 & 95.9 \\ 
1888.5 & 38.3 & 116.5 \\ 
1898.5 & 38.4 & 143.6 \\ 
1908.5 & 44.2 & 169.4 \\ 
1918.5 & 50.5 & 178.0 \\ 
1928.5 & 46.6 & 156.5 \\ 
1938.5 & 36.8 & 135.3 \\ 
1948.5 & 26.7 & 103.9 \\ 
1958.5 & 20.3 & 75.2 \\ 
1968.5 & 17.4 & 66.2 \\ 
1978.5 & 16.1 & 53.7 \\ 
1988.4 & 9.4 & 33.6 \\ 
1998.4 & 10.6 & 36.2 \\ 
2008.4 & 13.2 & 44.0 \\ 
2018.5 & 10.9 & 40.4 \\ 
2028.5 & 6.8 & 24.9 \\ 
2038.5 & 5.5 & 20.1 \\ 
2048.5 & 5.0 & 17.9 \\ 
2058.5 & 7.2 & 28.4 \\ 
2068.5 & 9.8 & 34.4 \\ 
2078.5 & 12.2 & 41.6 \\ 
2088.5 & 21.1 & 84.4 \\ 
2098.5 & 54.3 & 173.1 \\ 
2108.5 & 50.5 & 180.8 \\ 
2118.4 & 33.5 & 116.7 \\ 

\end{tabular} &
\begin{tabular}{ccc}
2128.4 & 34.6 & 120.4 \\ 
2138.4 & 48.4 & 164.4 \\ 
2148.4 & 53.2 & 172.8 \\ 
2158.3 & 46.3 & 149.2 \\ 
2168.3 & 32.9 & 108.4 \\ 
2178.3 & 27.4 & 91.1 \\ 
2188.2 & 30.5 & 98.9 \\ 
2198.2 & 28.9 & 94.0 \\ 
2208.2 & 35.1 & 113.1 \\ 
2218.2 & 39.9 & 134.8 \\ 
2228.2 & 48.0 & 154.1 \\ 
2238.3 & 53.3 & 157.6 \\ 
2248.3 & 54.8 & 180.5 \\ 
2258.3 & 57.9 & 179.7 \\ 
2268.2 & 61.1 & 185.2 \\ 
2278.2 & 76.8 & 227.2 \\ 
2288.1 & 79.0 & 220.8 \\ 
2298.0 & 75.6 & 223.9 \\ 
2308.0 & 75.3 & 256.6 \\ 
2318.0 & 78.0 & 264.4 \\ 
2328.0 & 59.4 & 193.7 \\ 
2338.0 & 51.8 & 159.1 \\ 
2348.0 & 32.9 & 118.7 \\ 
2357.9 & 23.9 & 86.2 \\ 
2367.8 & 28.6 & 104.0 \\ 
2377.8 & 32.5 & 119.7 \\ 
2387.8 & 44.5 & 140.6 \\ 
2397.9 & 56.9 & 203.5 \\ 
2408.0 & 57.5 & 201.8 \\ 
2418.0 & 54.3 & 178.2 \\ 
2428.1 & 61.3 & 206.0 \\ 
2438.1 & 68.8 & 205.2 \\ 
2448.2 & 77.6 & 232.1 \\ 
2458.2 & 79.8 & 259.4 \\ 
2468.3 & 89.4 & 276.1 \\ 
2478.4 & 80.7 & 244.3 \\ 
2488.4 & 56.7 & 185.2 \\ 

\end{tabular} \\
\end{tabular}
\caption{\label{res} Effective scattering length $1/b_e$ and absorption length $1/a$ at 400~nm vs.\ depth given at the $x,y$ coordinates corresponding to the center of IceCube array. This, together with the value of the shape parameter of the scattering function, $f_{\sf SL}=0.45$ constitues the ``SPICE Mie'' model. Additional parameters that this model depends on that were derived elsewhere are parameters $\alpha$, $\kappa$, $A$, and $B$ of the six-parameter ice model \cite{kurt}, and ice tilt map of \cite{ryan}. }
\end{center}\end{table}

\clearpage
\section*{Acknowledgement}
We acknowledge the support from the following agencies: U.S. National Science Foundation-Office of Polar Programs, U.S. National Science Foundation-Physics Division, University of Wisconsin Alumni Research Foundation, the Grid Laboratory Of Wisconsin (GLOW) grid infrastructure at the University of Wisconsin - Madison, the Open Science Grid (OSG) grid infrastructure; U.S. Department of Energy, and National Energy Research Scientific Computing Center, the Louisiana Optical Network Initiative (LONI) grid computing resources; National Science and Engineering Research Council of Canada; Swedish Research Council, Swedish Polar Research Secretariat, Swedish National Infrastructure for Computing (SNIC), and Knut and Alice Wallenberg Foundation, Sweden; German Ministry for Education and Research (BMBF), Deutsche Forschungsgemeinschaft (DFG), Helmholtz Alliance for Astroparticle Physics (HAP), Research Department of Plasmas with Complex Interactions (Bochum), Germany; Fund for Scientific Research (FNRS-FWO), FWO Odysseus programme, Flanders Institute to encourage scientific and technological research in industry (IWT), Belgian Federal Science Policy Office (Belspo); University of Oxford, United Kingdom; Marsden Fund, New Zealand; Australian Research Council; Japan Society for Promotion of Science (JSPS); the Swiss National Science Foundation (SNSF), Switzerland.

\bibliographystyle{nim/model1-num-names}

\end{document}